\documentclass[lettersize,journal]{IEEEtran}
\IEEEoverridecommandlockouts
\usepackage{cite}
\usepackage{amsmath,amssymb,amsfonts}
\usepackage{algorithmic}
\usepackage{graphicx}
\usepackage{textcomp}
\usepackage{xcolor}
\usepackage{multirow}
\usepackage[ruled,linesnumbered]{algorithm2e}
\usepackage{longtable}

\usepackage{array}  
\usepackage{makecell} 
\usepackage{amsthm}
\usepackage{tabularx}
\usepackage{booktabs}  

\usepackage{pifont}    

\newtheorem{proposition}{Proposition}

\usepackage{orcidlink}
\hypersetup{hidelinks}

\usepackage{makecell}

\usepackage{threeparttable} 
\usepackage{xcolor}

\def\BibTeX{{\rm B\kern-.05em{\sc i\kern-.025em b}\kern-.08em
    T\kern-.1667em\lower.7ex\hbox{E}\kern-.125emX}}

\begin{document}

\title{Unveiling Hidden Threats: Hierarchical Self-Similar Fractal Triggers for Stealthy Distributed Backdoor Attacks in Federated Learning\\

}

\author{Jian Wang\orcidlink{0000-0003-4537-0147}, Hong Shen\orcidlink{0000-0002-3663-6591}, and Chan-Tong Lam\orcidlink{0000-0003-0952-0961}
	\thanks{This work was supported in part by the Science and Technology Development Fund, Macao SAR under Grant 0033/2023/RIA1 and Macao Polytechnic University Research Gran RP/FCA-13/2022; in part by the Guangdong Provincial Department of Education under Grant 2024KTSCX133; and in part by the Queensland Department of Environment and Science Quantum Challenges 2032 Program under Grant Q2032001.}
	\thanks{J. Wang and Chan-Tong Lam are with the Faculty of Applied Sciences, Macao Polytechnic University, Macao SAR, China (e-mail: jian.wang@mpu.edu.mo).}
	\thanks{H. Shen is with the School of Engineering and Technology, Central Queensland University, Australia (e-mail: h.shen@cqu.edu.au). Corresponding author.}
	\thanks{J. Wang is also with the School of Cyberspace Security, Guangzhou University of Software, Guangzhou, China.}
}

\maketitle

\begin{abstract}
	The vulnerability of federated learning (FL) to distributed backdoor attacks, where malicious clients collaboratively implant hidden triggers into the global model, threatens the deployment of FL in real-world applications. 
	While spatial DBA fragments a completed trigger across clients, FracBA instead distributes the trigger generation process itself using fractal geometry. 
	By assigning different fractal iteration depths to malicious clients, FracBA preserves self-similarity across scales and enables coherent trigger accumulation during federated aggregation. This multi-scale construction is designed to align with hierarchical feature learning in neural networks.
	Across four evaluated datasets, FracBA achieves up to 97.4\% attack success rate; in an end-to-end comparison with spatial DBA, the full FracBA configuration requires fewer poisoned samples to reach the reported ASR threshold. 
	Against eight evaluated defenses, the local-estimate FracBA row in the defense table reports an average detection rate of 17.2\%, including 9.4\% under FLTrust. 
	These findings motivate further evaluation of FL backdoor defenses against hierarchically generated trigger constructions.
\end{abstract}


\section{Introduction}

\textbf{Background and Problem Setting.} 
Federated Learning (FL) enables multiple clients to collaboratively train machine learning models without sharing raw data, making it a foundational technique for privacy-sensitive domains such as healthcare, autonomous driving, and finance~\cite{mcmahan2017communication}. Recent improvements to FedAvg have been proposed to address various challenges. For instance, HyFDCA extends FedAvg to hybrid FL settings while maintaining comparable convergence rates~\cite{overman2024primal}.
Despite its advantages, the decentralized and partially trusted nature of FL also introduces unique security vulnerabilities.
Among them, backdoor attacks pose a particularly serious threat: a small fraction of malicious clients can inject hidden triggers during training, causing the global model to produce targeted misclassifications at inference time while maintaining high accuracy on benign inputs~\cite{byzantinegradientdescentHowBackdoorFederated2020,wangAttackTailsYes2020}.

In FL settings, backdoor attacks are more challenging to both execute and defend against than in centralized learning.
Attackers must operate under constrained participation, limited control over aggregation, and heterogeneous data distributions~\cite{karimireddy2020scaffold}, while defenders rely on indirect signals such as gradient statistics or activation patterns.
This interplay makes FL an especially relevant testbed for studying the fundamental limits of backdoor robustness and stealthiness.

Most existing backdoor attacks in FL rely on manually designed triggers, such as localized pixel patches, blended noise patterns, or small spatial perturbations~\cite{gu2019badnets}.
While these designs are effective in centralized learning, they face two critical limitations in federated environments.
First, simple triggers often introduce artificial structures that deviate from natural data statistics.
Such deviations manifest as abnormal spectral signatures or clustered activations, making them increasingly vulnerable to modern defenses based on spectral analysis, activation clustering, or trigger reverse engineering~\cite{tran2018spectral,wang2019neural,cao2020fltrust}.
Second, the effectiveness of these triggers tends to degrade under federated aggregation.
Since each client contributes only a small portion of the overall update, localized or dense perturbations are easily diluted, forcing attackers to increase poisoning rates or perturbation strength.
This trade-off between attack effectiveness and detectability becomes particularly pronounced in large-scale or highly heterogeneous FL systems.

To reduce per-client detectability, recent works have proposed \emph{distributed backdoor attacks}, where the trigger is fragmented across multiple malicious clients (DBA)~\cite{xieDBADISTRIBUTEDBACKDOOR2020}.
Most of these approaches follow a spatial decomposition paradigm, partitioning a complete trigger into geometric fragments assigned to different clients.
Although spatial decomposition reduces the visibility of individual updates, it disrupts global structural coherence.
As a result, partial triggers may interact inconsistently during aggregation, making attack success sensitive to client participation patterns and data heterogeneity.
More recent designs introduce semantic or dynamically transformed triggers to improve stealthiness~\cite{nguyen2022wanet}.
However, these methods typically rely on explicit adversarial optimization against specific defenses~\cite{doan2020februus}.
Such optimization-driven approaches are inherently brittle: improvements against one defense often come at the cost of vulnerability to others, and adapting to new defenses requires costly retraining or heuristic tuning.

\textbf{Key Insight: Distributing the Generation Process.} 
This work is motivated by a simple but fundamental observation:
existing distributed backdoor attacks focus on \emph{how to partition a trigger}, whereas the \emph{process by which the trigger is generated} remains largely unexplored.

We propose \textit{FracBA}, a novel backdoor attack framework that exploits the hierarchical generation process of fractal geometry.
Instead of spatially decomposing a complete trigger pattern, FracBA distributes different \emph{iteration depths} of a shared fractal construction, implemented via Iterated Function Systems (IFS), across malicious clients.
Early iterations generate coarse global structures, intermediate iterations introduce branching patterns, and later iterations refine fine-scale textures.
Each malicious client contributes a different stage of this generative process while sharing the same underlying transformation rules.

This shift from spatial partitioning to hierarchical generation yields several conceptual advantages.
Because all sub-triggers are generated using the same IFS transformations, structural self-similarity is preserved across clients, avoiding the scale inconsistency introduced by geometric cutting.
Sub-triggers produced at different iteration depths naturally correspond to complementary frequency bands.
During federated aggregation, these contributions align with the hierarchical feature extraction mechanisms of modern neural networks, enabling more stable and constructive reinforcement.

\textbf{Implications and Contributions.} 
Beyond aggregation behavior, fractal triggers also exhibit perceptual naturalness.
Their scale-invariant statistics resemble those of natural data, allowing them to blend into benign distributions and reducing susceptibility to frequency- or activation-based defenses.
These properties arise from intrinsic mathematical structure rather than explicit optimization against particular defenses.

FracBA departs from existing distributed backdoor paradigms by distributing the trigger generation process itself rather than partitioning a completed pattern.

This work makes the following contributions:

\begin{itemize}
	\item \textbf{Fractal Generation Paradigm.}
	By distributing the trigger's \emph{generation process} rather than its spatial fragments, FracBA is designed to preserve cross-client structural consistency across hierarchical levels; a controlled comparison with spatial DBA shows a lower poisoned-sample requirement at the reported ASR threshold, without an asymptotic guarantee.
	
	\item \textbf{Complete FracBA Scheme.}
	The scheme covers IFS generation and parameter selection, hierarchical layer assignment, three-stage dynamic angular perturbation ($5^\circ/15^\circ/5^\circ$), and strategic poisoning based on local information, balancing attack effectiveness, stealthiness, and aggregation robustness.
	
	\item \textbf{Systematic Evaluation.}
	Through evaluation on four datasets under IID and Non-IID settings, the reported defense settings, and sparse participation, FracBA achieves high ASR and low reported detection rates; its poisoned-sample efficiency is supported specifically by the controlled comparison with spatial DBA.
\end{itemize}

\section{Related Work}
\label{sec:related}

\subsection{Backdoor Attacks in Federated Learning}

Backdoor attacks have been extensively studied in centralized learning settings, where attackers inject triggers into training data to induce targeted misclassification at inference time~\cite{liu2018trojaning}.
In FL, the attack surface is fundamentally altered by distributed training, partial client participation, and server-side aggregation.
Early studies demonstrate that even a small fraction of malicious clients can successfully implant backdoors by manipulating local updates.

However, all these approaches rely on triggers with fixed spatial structures 
that can be partitioned or distributed but not fundamentally decomposed in 
their generation process. Recent surveys~\cite{nguyen2024backdoor,feng2025survey,li2025backdoor} 
systematically categorize these attacks, confirming that existing methods 
focus on trigger partitioning rather than distributed generation.
Recent attacks further improve stealthiness through parameter-subspace poisoning or layer-wise gradient alignment~\cite{guo2026pill,yang2025stealthy}, while defense-side work has explored detection based on latent-space trigger collision~\cite{xiong2026trigger}. On the defense side, beyond statistical filtering and robust aggregation~\cite{cao2020fltrust,nguyen2022flame}, recent mitigation frameworks include DeepSight~\cite{rieger2022deepsight}, which inspects client models, and FLIP~\cite{zhang2023flip}, which provides provable guarantees.

\subsection{Distributed Backdoor Attacks via Spatial Decomposition}

To reduce per-client detectability, recent works propose \emph{distributed backdoor attacks} that split the trigger across multiple malicious clients.
DBA pioneers this paradigm by spatially decomposing a complete trigger into geometric fragments, assigning each fragment to a different client.
This design reduces the visual saliency of individual triggers and improves resistance to simple per-client anomaly detection.

Subsequent works extend this idea.
Attack-Tails~\cite{wangAttackTailsYes2020} adaptively adjusts poisoning intensity over training rounds to compensate for aggregation dilution, while maintaining the same spatial partitioning principle.
Gong et al.~\cite{gongCoordinatedBackdoorAttacks2022} further coordinate multiple local triggers using model-dependent patterns to enhance post-aggregation attack effectiveness.
Despite their effectiveness, spatial decomposition approaches exhibit inherent limitations.
Because each client contributes only a spatially localized fragment, the global trigger must be reconstructed implicitly through aggregation.
This reconstruction is sensitive to client sampling, data heterogeneity, and model architecture, often requiring increased poisoning rates or stronger perturbations to maintain attack success.

Spatial partitioning disrupts global structural coherence.
Fragments lack scale consistency and may activate heterogeneous feature subsets during training, leading to unstable reinforcement and reduced persistence under long-term federated optimization.

\subsection{Semantic and Optimization-Driven Trigger Designs}

Beyond geometric partitioning, several works explore alternative strategies to improve stealthiness.
Semantic triggers embed naturally occurring objects or patterns into training data, aiming to avoid conspicuous artificial artifacts~\cite{byzantinegradientdescentHowBackdoorFederated2020}.
While visually plausible, such triggers rely heavily on task-specific semantics and often generalize poorly across datasets or domains.

Other approaches introduce dynamically transformed or invisible triggers.
WaNet~\cite{nguyen2022wanet} employs spatial warping to generate imperceptible perturbations, while ISSBA~\cite{li2021invisible} uses encoder--decoder architectures to hide triggers in latent representations. 

These methods typically rely on explicit adversarial optimization against particular defenses.
As a result, their effectiveness can be brittle: improvements against one detection mechanism may degrade robustness to others, and adapting to new defenses often requires substantial retraining or heuristic tuning.

\subsection{Limitations of Existing Paradigms}

Despite significant progress, existing backdoor attack paradigms in FL share common limitations.
Patch-based triggers are prone to detection due to their artificial structure.
Spatially decomposed triggers reduce per-client visibility but sacrifice global coherence, making them sensitive to aggregation dynamics.
Semantic and optimization-driven triggers improve perceptual stealthiness but lack principled guarantees under aggregation and transformation.

All existing distributed backdoor approaches focus on \emph{partitioning a completed trigger pattern}.
The process by which the trigger is generated remains centralized and monolithic, and is only fragmented after construction.
This design choice implicitly assumes that spatial fragments can be reliably recombined during federated aggregation, an assumption that often fails in practice.

\subsection{Hierarchical Trigger Generation via Fractals}

Our work departs from prior paradigms by distributing the \emph{generation process} of the trigger rather than its spatial fragments.
FracBA builds on the hierarchical construction of fractal geometry through IFS, assigning different iteration depths to different malicious clients.
Each client generates a structurally consistent sub-trigger derived from the same transformation rules, but operating at a distinct scale.

This hierarchical decomposition preserves self-similarity across clients and produces sub-triggers that correspond to complementary frequency components.
Unlike spatial partitioning, hierarchical generation maintains scale coherence and aligns naturally with the multi-layer feature extraction mechanisms of modern neural networks.
As a result, backdoor signals from different clients can be aggregated more constructively and with improved stability.

Overall, existing distributed backdoor attacks differ primarily in how triggers are constructed and distributed across clients, but they all rely on partitioning a completed pattern.
In contrast, FracBA distributes the trigger generation process itself, leading to fundamentally different aggregation behavior.

\section{Threat Model}

We consider a standard cross-device FL setting with a central server and a population of $N$ clients. The server coordinates training over multiple communication rounds by periodically selecting a subset of clients, distributing the current global model, and aggregating their locally updated models. We assume the server follows the prescribed aggregation protocol (e.g., FedAvg~\cite{mcmahan2017communication}) and does not behave adversarially.

\subsection{Attacker Capabilities}
\label{sec:attacker_capability}

The attacker controls a subset of malicious clients and participates in the FL process as a legitimate client. These malicious clients follow the standard training and communication protocol, but manipulate their local training data to introduce a backdoor. We assume the attacker cannot modify the server-side aggregation algorithm, access other clients' private data, or interfere with client selection beyond normal participation.

Consistent with prior work, we assume that only a minority of clients are malicious, and that malicious clients may be sampled intermittently across training rounds due to random client selection. The attacker is therefore constrained to operate in a distributed and stochastic environment, where malicious updates must compete with benign updates during aggregation.

In addition, the attacker can only exploit locally available information: each malicious client uses a moving average of its own trigger-free update norms as a local proxy for a benign-update scale, without access to server-side or other clients' statistics. This proxy is not calibrated to the population of benign clients and need not represent that population under Non-IID data.

\subsection{Attacker Goals}

The attacker's goal is to implant a backdoor into the global model such that inputs containing a specific trigger pattern are misclassified into a target label, while the model's performance on clean inputs remains largely unaffected. We focus on persistent backdoor attacks, where the malicious behavior must survive aggregation across multiple training rounds rather than being injected in a single round.

The attacker does not seek to maximize attack strength at any cost. Excessive deviation from benign updates may raise suspicion or be filtered by robust aggregation mechanisms. As such, the attacker aims to balance attack effectiveness with stealthiness and persistence under aggregation noise.

\subsection{Attack Constraints}
We impose the following constraints to reflect realistic FL deployments.
First, the attacker controls only a limited fraction of clients and cannot guarantee participation in every training round.
Second, the attacker has no access to server-side information or defenses beyond what can be inferred from the global model.
Third, the backdoor trigger should remain perceptually natural, without trivially altering the semantic content of inputs or introducing obvious distributional artifacts; perceptual naturalness is treated as a secondary goal rather than a hard constraint.
Finally, the injected backdoor must persist despite dilution by benign updates, stochastic client sampling, and potential noise introduced by the system.

\subsection{Defender Model}
We assume a non-adaptive but security-aware defender. The server may employ common mitigation strategies proposed in the literature, such as robust aggregation~\cite{sunCanYouReally2019}, anomaly detection on client updates, or post-hoc model inspection~\cite{chen2019deepinspect,liu2023detecting}. However, the defender does not have prior knowledge of the specific trigger pattern or attack strategy used by the attacker.

We do not assume that defenses are ineffective by default. Instead, our analysis and experiments examine how the proposed attack behaves under different system conditions and defensive pressures, and where its effectiveness degrades.

\subsection{Relation to Prior Threat Models}
Our threat model aligns with standard FL backdoor attack settings, while emphasizing distributed and persistent attacks under realistic constraints. Unlike attacks that rely on aggressive weight manipulation or direct model replacement, we focus on trigger-driven attacks that must remain compatible with benign training dynamics and aggregation mechanisms. This design choice allows us to study the structural and statistical conditions under which such attacks succeed or fail.

\section{Method}
\label{sec:method}

\subsection{Overview}
\begin{figure*}[t]
\centering
\includegraphics[width=0.9\textwidth]{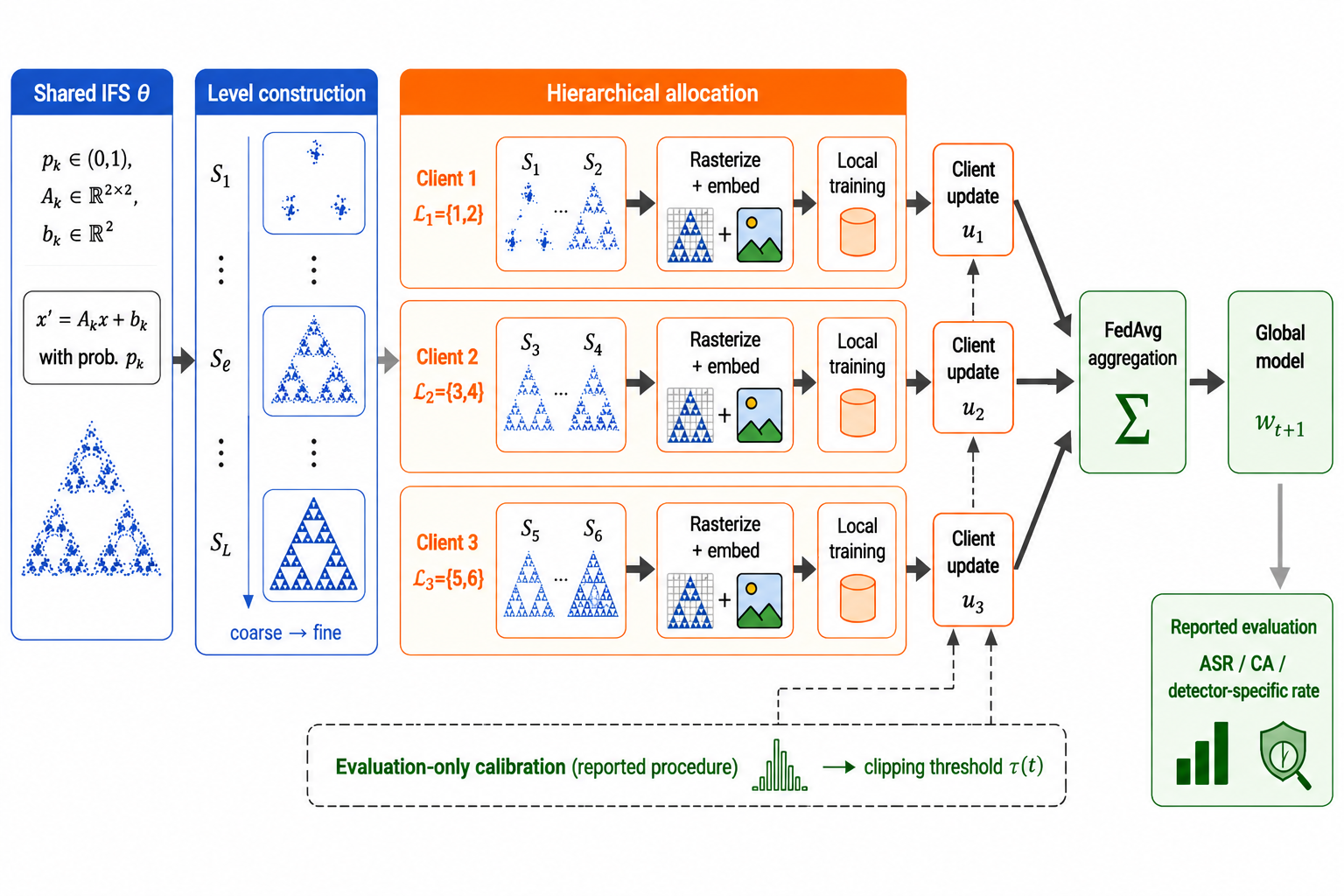}
\caption{Overview of the FracBA framework. Fractal triggers are generated by IFS and distributed across malicious clients along the iteration hierarchy; each client embeds its sub-trigger into local samples, applies dynamic angular perturbation and strategic scaling/clipping, and submits the update for FedAvg aggregation into the global model.}
\label{fig:overview}
\end{figure*}

FracBA exploits fractal patterns' iterative construction properties to generate visually natural yet statistically evasive triggers for backdoor attacks in FL. Our approach consists of four key components: (1) \textit{Fractal trigger generation and iterative decomposition} using IFS to create client-specific sub-triggers through hierarchical layer assignment, where each client contributes features at different spatial scales; (2) \textit{Trigger embedding strategy} to integrate sub-triggers into clean samples while preserving image semantics through adaptive blending; (3) \textit{Dynamic angular perturbation mechanism} to further disrupt frequency-domain signatures across training phases while maintaining attack efficacy via structural invariance; and (4) \textit{Strategic poisoning in FL} with gradient masking and adaptive scaling tailored to aggregation dynamics.
Figure~\ref{fig:overview} provides an overview of the FracBA framework.

\subsection{Fractal Trigger Generation}
\label{sec:fractal_generation}

\textbf{Motivation.} Traditional triggers exhibit simple statistical signatures detectable by spectral analysis~\cite{tran2018spectral}. Fractals offer three advantages: natural appearance, distributed spectral energy, and multi-scale robustness. In FL scenarios, distributing trigger components across clients based on their \textit{iterative construction process} enhances both security and stealthiness. We note that visual imperceptibility is not a strict requirement for backdoor injection in FL, since the attacker already operates at the training stage. In this work, perceptual naturalness is considered as a secondary property that correlates with statistical similarity to benign data, which in turn affects the effectiveness of existing detection-based defenses.

\textbf{IFS-based Generation.} We generate fractal triggers using Iterated Function Systems, defined by a set of affine transformations $\{T_1, \ldots, T_K\}$ with probabilities $\{p_1, \ldots, p_K\}$. Each transformation $T_k(\mathbf{x}) = \mathbf{A}_k \mathbf{x} + \mathbf{b}_k$ maps points in 2D space, where $\mathbf{A}_k \in \mathbb{R}^{2 \times 2}$ controls scaling/rotation and $\mathbf{b}_k \in \mathbb{R}^2$ controls translation. Starting from an initial point $\mathbf{x}_0$, we iteratively apply transformations to generate points at increasing depth levels:
\vspace{-0.1cm}
\begin{equation}
	\mathbf{x}_{t+1} = T_k(\mathbf{x}_t), \quad k \sim \text{Categorical}(p_1, \ldots, p_K)
\end{equation}
\vspace{-0.1cm}
Here, $t$ denotes the iteration index; after $M$ iterations, the point cloud $\{\mathbf{x}_1, \ldots, \mathbf{x}_M\}$ forms a finite approximation of a fractal attractor with self-similar structure across scales.

\textbf{Adaptive Parameter Selection.} To ensure dataset compatibility, we optimize IFS parameters 
\begin{equation}
	\Theta = \{\mathbf{A}_k, \mathbf{b}_k, p_k\}_{k=1}^K
\end{equation}

via gradient descent on a small validation set. The objective balances attack success and imperceptibility:
\vspace{-0.1cm}
\begin{equation}
	\min_{\Theta} \mathcal{L}(\Theta) = \mathcal{L}_{\text{attack}}(\Theta) + \lambda \mathcal{L}_{\text{percep}}(\Theta)
\end{equation}
\vspace{-0.1cm}
where 
\begin{equation}
	\mathcal{L}_{\text{attack}} = -\log P(y_t | \mathbf{x} + \delta_\Theta)
\end{equation}
measures misclassification to target label 
$y_t$,
$\delta_\Theta$ denotes the trigger generated from parameters $\Theta$, and 
\begin{equation}
	\mathcal{L}_{\text{percep}} = \|\delta_\Theta\|_2^2 + \text{LPIPS}(\mathbf{x}, \mathbf{x} + \delta_\Theta)
\end{equation}

\textbf{Iterative Decomposition via Hierarchical Layer Assignment.} To enhance stealthiness in FL, we decompose the fractal generation process itself rather than partitioning the final pattern. This approach uses the inherent multi-scale structure of IFS: early iterations produce coarse-scale features, while later iterations add fine-scale details.

Given optimized IFS parameters $\Theta = \{T_1, \ldots, T_K\}$, we organize the iteration process into $L$ hierarchical levels and allocate $m_\ell$ sampled points to level $\ell$, where $\sum_{\ell=1}^L m_\ell=M$. Starting from $\mathbf{x}_0$, each level $\ell$ generates an $m_\ell$-element multiset through independently sampled categorical addresses:
\vspace{-0.1cm}
\begin{equation}
	\begin{aligned}
	S_\ell &= \biguplus_{r=1}^{m_\ell}\{\mathbf{x}_{r,\ell}\},\\
	\mathbf{x}_{r,\ell} &= T_{k_{r,\ell}} \circ \cdots \circ T_{k_{r,1}}(\mathbf{x}_0),\\
	k_{r,j} &\overset{\mathrm{i.i.d.}}{\sim} \text{Categorical}(p_1, \ldots, p_K).
	\end{aligned}
\end{equation}
\vspace{-0.1cm}
Thus, $S_\ell$ represents sampled addresses rather than all $K^\ell$ transformation compositions. In round $t$, let $\mathcal{A}^{(t)}=\mathcal{S}^{(t)}\cap\mathcal{M}$ denote the active malicious clients selected by the server. If $\mathcal{A}^{(t)}=\emptyset$, no levels are grouped and no attack is performed in that round. Otherwise, we group the $L$ levels into $G=\min(L,|\mathcal{A}^{(t)}|)$ ordered level groups $\{\mathcal{B}_g\}_{g=1}^G$, and assign active client $i$ a subset of levels $\mathcal{L}_i^{(t)} \subset \{1,\ldots,L\}$ such that:
\vspace{-0.1cm}
\begin{equation}
	\bigcup_{i \in \mathcal{A}^{(t)}} \mathcal{L}_i^{(t)} = \{1,\ldots,L\}, \quad
	\mathcal{L}_i^{(t)} \cap \mathcal{L}_j^{(t)} = \emptyset \ (i\ne j).
\end{equation}
\vspace{-0.1cm}

Client $i$ generates its sub-trigger by rasterizing only points from its assigned levels:
\vspace{-0.1cm}
\begin{equation}
	\delta_i^{(t)} = \text{Rasterize}\left( \biguplus_{\ell \in \mathcal{L}_i^{(t)}} S_\ell \right), \quad \text{normalized to } [0, \epsilon]
\end{equation}
\vspace{-0.1cm}
The rasterization operation maps point clouds to a discrete image grid via Gaussian kernel density accumulation followed by normalization.

\textbf{Example Configuration.} For $L=6$ levels, $K=4$ transformations, and $|\mathcal{A}^{(t)}|=3$ active malicious clients in a given round:
\begin{itemize}
	\item Active client 1: $\mathcal{L}_1^{(t)} = \{1,2\}$, so $\delta_1^{(t)}$ contains \textit{coarse-scale features} (a sparse point cloud defining global structure)
	\item Active client 2: $\mathcal{L}_2^{(t)} = \{3,4\}$, so $\delta_2^{(t)}$ contains \textit{mid-scale features} (intermediate-density bridging structures)
	\item Active client 3: $\mathcal{L}_3^{(t)} = \{5,6\}$, so $\delta_3^{(t)}$ contains \textit{fine-scale features} (dense textures and high-frequency details)
\end{itemize}

This hierarchical decomposition provides three critical advantages:

\textbf{(1) Preserved Fractal Properties:} Each level $S_\ell$ is generated through shared IFS transformations and is designed to retain related multi-scale structure in every sub-trigger $\delta_i$. The observed interval $D_i \in [D-0.15, D+0.15]$ is an empirical measurement for the evaluated finite-resolution triggers, not a guarantee for arbitrary IFS parameters or samples.

\textbf{(2) Multi-Scale Complementarity:} Different clients contribute features at complementary spatial scales. During federated aggregation, model gradients from $\delta_1$ (coarse), $\delta_2$ (medium), and $\delta_3$ (fine) combine to form a complete multi-scale backdoor representation. This mirrors how IFS iteratively refines fractal patterns: early iterations establish structure, while later iterations add detail.

\textbf{(3) Enhanced Stealthiness:} Individual sub-triggers appear as sparse, irregular point clouds rather than structured patterns. For example, $\delta_1$ with only $|S_1| + |S_2| \approx 3,300$ points (for $M=10^4, L=6$) exhibits lower spatial autocorrelation and weaker frequency peaks compared to the full pattern, reducing detectability via spectral analysis or activation clustering defenses.

\subsection{Trigger Embedding Strategy}
\label{sec:trigger_embedding}

After generating level-specific sub-triggers, each malicious client embeds its assigned pattern into local training samples. Given a sub-trigger $\delta_i \in [0, \epsilon]^{H \times W}$ (where $\epsilon$ controls the overall trigger strength) and a clean sample $\mathbf{x} \in [0, 1]^{H \times W \times 3}$, both $\delta_i$ and the spatial mask $\mathbf{M}$ are broadcast across the three image channels, and the poisoned sample is constructed as:
\begin{equation}
	\label{eq:trigger_embedding}
	\mathbf{x}_{\text{poisoned}}^{(i)} = \text{clip}(\mathbf{x} + \alpha_{\text{blend}} (\delta_i \odot \mathbf{M}), 0, 1)
\end{equation}
where $\alpha_{\text{blend}} \in [0.15, 0.3]$ controls the blending intensity (we use $\alpha_{\text{blend}}=0.2$ in our experiments), and $\mathbf{M} \in \{0,1\}^{H \times W}$ is a binary spatial mask defining the trigger region. Following common practice in backdoor attacks, we use a localized mask region at the bottom-right corner of the image: $10\times10$ on CIFAR-10/GTSRB (about 9.8\% of the image), $20\times20$ on Tiny-ImageNet (about 9.8\%), and $56\times56$ on ImageNet-100 (about 6.25\%), to maintain trigger locality while preserving the semantic content of the original image. The clipping operation ensures pixel values remain in the valid range $[0, 1]$ after perturbation.

This localized embedding strategy offers two key advantages: (1) it preserves the fractal's intrinsic spectral properties within the trigger region, making sub-triggers statistically indistinguishable from natural textures under frequency-domain analysis; (2) it maintains natural image statistics in the majority of the image, further enhancing inconspicuousness against automated detection methods. Because different clients embed sub-triggers at different scales, the poisoned datasets across clients exhibit structural diversity in the frequency domain, reducing the risk of detection via cross-client spectral clustering or activation pattern analysis.

\subsection{Dynamic Angular Perturbation Mechanism}
\label{sec:dynamic_perturbation}

Building upon the iterative decomposition architecture in Section~\ref{sec:fractal_generation}, we introduce a second layer of defense evasion through \textit{dynamic angular perturbation}. While level-based decomposition provides multi-scale diversity, IFS-generated patterns still exhibit detectable spectral regularities (e.g., periodic angular components in Fourier space arising from repeated transformation applications). To disrupt these signatures, each malicious client applies time-varying rotation to its sub-trigger before embedding.

\textbf{Perturbation Strategy.} At round $t$, client $i$ generates a perturbed version of its sub-trigger via rotation:
\vspace{-0.1cm}
\begin{equation}
	\tilde{\delta}_i^{(t)} = \text{Rotate}(\delta_i, \Delta\theta_i^{(t)})
	\label{eq:rotation}
\end{equation}

\vspace{-0.1cm}
where $\Delta\theta_i^{(t)} \in [-\Delta\theta_{\max}, +\Delta\theta_{\max}]$ is sampled uniformly for each client and round. Rotation is performed around the center of the trigger region.

The perturbation magnitude follows a three-stage schedule aligned with federated training dynamics:

\begin{itemize}
	\item \textbf{Stage 1 (Establishment, first 30\% of rounds):} Small perturbation ($\Delta\theta_{\max} = 5^\circ$) allows the global model to establish initial backdoor correlations with minimal geometric variance across clients.
	\item \textbf{Stage 2 (Evasion, middle 40\% of rounds):} Large perturbation ($\Delta\theta_{\max} = 15^\circ$) disrupts spectral regularities to evade frequency-domain defenses during their typical detection window.
	\item \textbf{Stage 3 (Consolidation, last 30\% of rounds):} Reduced perturbation ($\Delta\theta_{\max} = 5^\circ$) stabilizes backdoor convergence and supports high attack success at deployment.
\end{itemize}

\textbf{Preservation of Attack Efficacy Under Perturbation.} A critical question arises: why does angular perturbation (up to $\pm 15^\circ$) combined with level-based decomposition not disrupt the backdoor? The answer lies in two complementary mechanisms:

\textit{(1) Scale-Invariant Fractal Properties:} Each sub-trigger $\delta_i$ generated from level subset $\mathcal{L}_i$ retains the intrinsic statistical properties of the IFS. Rotation is an orthogonal transformation that preserves Euclidean distances and local point density; thus, the power spectral density and fractal dimension of $\text{Rotate}(\delta_i, \theta)$ remain nearly identical to those of $\delta_i$. During federated aggregation, gradients from rotated sub-triggers at different scales still share overlapping frequency components, allowing the global model to learn a unified multi-scale backdoor representation.

\textit{(2) Structural Invariance Under Rotation:} While rotation angles vary across clients and rounds, the underlying IFS contractive maps $\{T_k\}_{k=1}^K$ defining each level remain fixed: their scaling factors, translation vectors, and compositional structure are unchanged. During aggregation, model gradients converge to \textit{topological invariants} (connectivity patterns, spectral density distribution, fractal dimension) rather than exact pixel-level geometric configurations. These structural properties are robust to rotation, allowing the global model to learn a generalized backdoor manifold encompassing all perturbed variants $\{\tilde{\delta}_i^{(t)}\}$ across scales.

\textit{Gradient Accumulation Dynamics:} Stage 1's small perturbations establish an initial basin of attraction in the loss landscape. Stage 2's larger perturbations explore this basin's boundaries without escaping it, thanks to the cumulative effect of previous rounds' gradients and the inherent smoothness of fractal patterns under rotation. Stage 3 consolidates learned features, encouraging convergence toward a stable multi-scale backdoor representation.

\subsection{Strategic Poisoning in Federated Learning}
\label{sec:strategic_poisoning}

\textbf{Poisoned Dataset Construction.} Each malicious client $i \in \mathcal{M}$ constructs a poisoned dataset $\mathcal{D}_i^p$ by selecting a random fraction $\alpha$ (typically $\alpha=0.3$) of its local data and replacing them with triggered versions:
\vspace{-0.1cm}

\begin{equation}
	\mathcal{D}_i^p =
	\left\{
	(\mathbf{x}_{j,\text{poisoned}}^{(i,t)}, y_t)
	\right\}_{j \in \mathcal{I}_i}
\end{equation}
$$\mathbf{x}_{j,\text{poisoned}}^{(i,t)}
=
\text{clip}\left(
\mathbf{x}_j^{(i)}
+
\alpha_{\text{blend}}
\tilde{\delta}_i^{(t)} \odot \mathbf{M},
0,1
\right)
$$
\vspace{-0.1cm}
where $\tilde{\delta}_i^{(t)}$ is the client's perturbed sub-trigger at round $t$ (from Eq.~\ref{eq:rotation}), and $y_t$ is the target label. The remaining $(1-\alpha)|\mathcal{D}_i|$ samples remain clean to maintain model utility on the main task.

\textbf{Model Update Crafting.} To amplify backdoor injection while evading detection, we employ scaled local training:
\vspace{-0.1cm}
\begin{equation}
	\mathbf{w}_i^{t+1} = \mathbf{w}_t - \eta \nabla \mathcal{L}(\mathbf{w}_t; \mathcal{D}_i^p \cup \mathcal{D}_i^{\text{clean}})
\end{equation}
\vspace{-0.1cm}
followed by strategic scaling: $$\tilde{\mathbf{w}}_i^{t+1} = \mathbf{w}_t + \gamma(\mathbf{w}_i^{t+1} - \mathbf{w}_t)$$, where $\gamma > 1$ boosts the update magnitude. To balance influence and stealthiness, we fix $\gamma=2.0$ in our experiments (see Table~\ref{tab:fl_config}); Section~\ref{sec:ablation} reports a sensitivity analysis showing that this value balances attack success and stealthiness.

\textbf{Defense Evasion.} To evade norm-based defenses (e.g., FLTrust, FoolsGold), we clip updates if $\|\tilde{\mathbf{w}}_i^{t+1} - \mathbf{w}_t\|_2$ exceeds a dynamically computed threshold:
\vspace{-0.1cm}
\begin{equation}
	\begin{aligned}
	\hat{\tau}^{(t)} &= \hat{\mu}_b^{(t)} + 2\hat{\sigma}_b^{(t)},\\
	d_t(\mathbf{v}) &= \|\mathbf{v}-\mathbf{w}_t\|_2,\\
	s_t(\mathbf{v}) &=
	\begin{cases}
	1, & d_t(\mathbf{v})\leq\hat{\tau}^{(t)},\\
	\hat{\tau}^{(t)}/d_t(\mathbf{v}), & \text{otherwise},
	\end{cases}\\
	\operatorname{Clip}_t(\mathbf{v})
	&= \mathbf{w}_t+s_t(\mathbf{v})(\mathbf{v}-\mathbf{w}_t),\\
	\mathbf{u}_i^{t+1}
	&=\operatorname{Clip}_t(\tilde{\mathbf{w}}_i^{t+1}).
	\end{aligned}
\label{eq:clip}
\end{equation}
\vspace{-0.1cm}
where $\hat{\mu}_b^{(t)}$ and $\hat{\sigma}_b^{(t)}$ denote the baseline median and standard deviation estimated by the malicious client from a moving average of its own trigger-free update norms, consistent with the attacker capability assumed in Section~\ref{sec:attacker_capability}; we report this local-estimate variant as the main result (see the note under Table~\ref{tab:defense_resistance}). In an oracle version used only as an explanatory reference, these quantities correspond to the median and standard deviation of benign client update norms. The operator leaves models within the threshold unchanged and radially rescales models outside it. For spectral defenses~\cite{tran2018spectral}, the combination of level-based decomposition and angular perturbation is evaluated empirically in Section~\ref{sec:visualization}.

\begin{algorithm}[t]
	\small
	\caption{FracBA: Hierarchical Fractal Backdoor Attack}
	\label{alg:fracba}
	\begin{algorithmic}[1]
		\STATE \textbf{Input:} Malicious client pool $\mathcal{M}$, target label $y_t$, IFS parameters $\Theta$, levels $L$, scaling $\gamma$
		\STATE \textbf{Initialize:} Optimized IFS parameters $\Theta$
		\FOR{each FL round $t = 1, 2, \ldots, T$}
		\STATE Server broadcasts global model $\mathbf{w}_t$
		\STATE Active malicious set: $\mathcal{A}^{(t)} \leftarrow \mathcal{S}^{(t)}\cap\mathcal{M}$
		\IF{$\mathcal{A}^{(t)}=\emptyset$}
		\STATE No level grouping or attack is performed; selected benign clients proceed normally
		\ENDIF
		\STATE Partition levels among $\mathcal{A}^{(t)}$: $\{\mathcal{L}_i^{(t)}\}_{i\in\mathcal{A}^{(t)}}$
		\STATE Determine perturbation magnitude $\Delta\theta_{\max}^{(t)}$ based on the current training stage \hfill 
		\FOR{each malicious client $i \in \mathcal{A}^{(t)}$}
		\STATE Generate points for assigned levels: $S_\ell \leftarrow \text{IFS}^{\ell}(\Theta, \mathbf{x}_0)$ for $\ell \in \mathcal{L}_i^{(t)}$ \hfill
		\STATE Rasterize sub-trigger: $\delta_i^{(t)} \leftarrow \text{Rasterize}(\bigcup_{\ell \in \mathcal{L}_i^{(t)}} S_\ell)$ \hfill
		\STATE Sample rotation: $\Delta\theta_i^{(t)} \sim \text{Uniform}(-\Delta\theta_{\max}^{(t)}, +\Delta\theta_{\max}^{(t)})$
		\STATE Perturb sub-trigger: $\tilde{\delta}_i^{(t)} \leftarrow \text{Rotate}(\delta_i^{(t)}, \Delta\theta_i^{(t)})$ \hfill 
		\STATE Embed trigger: $\mathbf{x}_{\text{poisoned}}^{(i)} \leftarrow \text{clip}(\mathbf{x} + \alpha_{\text{blend}}(\tilde{\delta}_i^{(t)} \odot \mathbf{M}), 0, 1)$ \hfill 
		\STATE Construct poisoned dataset: $\mathcal{D}_i^p \leftarrow \{(\mathbf{x}_{\text{poisoned}}^{(i)}, y_t)\}$ \hfill 
		\STATE Local training: $\mathbf{w}_i^{t+1} \leftarrow \mathbf{w}_t - \eta \nabla \mathcal{L}(\mathbf{w}_t; \mathcal{D}_i^p \cup \mathcal{D}_i^{\text{clean}})$ \hfill 
		\STATE Scale update: $\tilde{\mathbf{w}}_i^{t+1} \leftarrow \mathbf{w}_t + \gamma(\mathbf{w}_i^{t+1} - \mathbf{w}_t)$
		\STATE Apply adaptive clipping via Eq.~\eqref{eq:clip} if $\|\tilde{\mathbf{w}}_i^{t+1} - \mathbf{w}_t\|_2 > \hat{\tau}^{(t)}$
		\STATE Let $\mathbf{u}_i^{t+1}$ denote the submitted local model after scaling and Eq.~\eqref{eq:clip}
		\STATE Submit $\mathbf{u}_i^{t+1}$ to server
		\ENDFOR
		\STATE Benign selected clients set $\mathbf{u}_i^{t+1}\leftarrow\mathbf{w}_i^{t+1}$
		\STATE Set $q_i^{(t)}\leftarrow |\mathcal{D}_i|/\sum_{j\in\mathcal{S}^{(t)}}|\mathcal{D}_j|$
		\STATE Aggregate: $\mathbf{w}_{t+1}\leftarrow\sum_{i\in\mathcal{S}^{(t)}}q_i^{(t)}\mathbf{u}_i^{t+1}$
		\ENDFOR
		\STATE \textbf{Output:} Backdoored global model $\mathbf{w}_T$
	\end{algorithmic}
\end{algorithm}

{\sloppy Algorithm~\ref{alg:fracba} summarizes the complete FracBA procedure. The key innovation is the synergistic combination of three mechanisms: (1) iterative level-based decomposition that distributes multi-scale fractal features across clients while preserving self-similarity, (2) dynamic angular perturbation for frequency-domain evasion through rotation-invariant topological properties, and (3) adaptive gradient masking for aggregation-level stealthiness. Together, these components enable triggers that are simultaneously natural (fractal geometry), distributed (hierarchical multi-scale construction), evasive (multi-layer obfuscation), and effective (preserved attack success through scale-invariant properties) across diverse FL scenarios.}

\section{Theoretical Analysis}
\label{sec:theory}

This section provides limited analytical mechanisms rather than a general convergence or sample-complexity theory for FracBA. The practical triggers are finite-resolution, perturbed rasterizations, so exact self-similar IFS results do not directly characterize the trained FL model. We isolate two deterministic stability statements and use qualitative or heuristic language for the remaining mechanisms.

\subsection{Qualitative Sample-Efficiency Mechanism}

Hierarchical generation uses the same IFS maps across levels, so client-specific sub-triggers are designed to retain related cross-client structure instead of being disjoint spatial pieces. This shared construction can support more coherent aggregation when the relevant malicious clients participate. In the controlled comparison of Section~\ref{sec:theory_validation}, the hierarchical construction requires fewer poisoned samples than spatial DBA to reach the reported ASR threshold. This observation is specific to that protocol and does not imply an asymptotic sample-complexity guarantee, a fixed correlation bound, or monotonic improvement as the number of malicious clients grows.

\subsection{Detection-Resistance Analysis}

\subsubsection{Gradient-Based Defenses}

Gradient norm filters (e.g., FLTrust~\cite{cao2020fltrust}, FLAME~\cite{nguyen2022flame}) use update magnitude as one detection signal. The following proposition states only the deterministic consequence of the explicit clipping operator in Eq.~\eqref{eq:clip}; it is not a detection-probability guarantee.

\begin{proposition}[Post-Clipping Norm Constraint]
	\label{thm:gradient}
	Let $\mathbf{u}_i^{t+1}=\operatorname{Clip}_t(\tilde{\mathbf{w}}_i^{t+1})$ be the submitted local model defined in Eq.~\eqref{eq:clip}, with $\hat{\tau}^{(t)}=\hat{\mu}_b^{(t)}+2\hat{\sigma}_b^{(t)}\geq0$. Then
	\begin{equation}
		\|\mathbf{u}_i^{t+1}-\mathbf{w}_t\|_2 \leq \hat{\tau}^{(t)}.
	\end{equation}
\end{proposition}
When the local estimate is close to the benign update scale, this constraint can weaken signals based solely on unusually large norms; the closeness of that estimate and the resulting detection behavior remain empirical questions.

\subsubsection{Spectral Defenses}
Let $\boldsymbol{\Sigma}\succ0$ be a fixed covariance matrix and define the point-to-point Mahalanobis metric
\begin{equation}
	D_{\boldsymbol{\Sigma}}(\mathbf{a},\mathbf{b})=\sqrt{(\mathbf{a}-\mathbf{b})^\top\boldsymbol{\Sigma}^{-1}(\mathbf{a}-\mathbf{b})}.
\end{equation}

\begin{proposition}[Feature Stability Under a Fixed Mahalanobis Metric]
	\label{thm:spectral}
	If the feature extractor $\mathbf{h}$ is $L_\phi$-Lipschitz in the Euclidean norm, then for any input $\mathbf{x}$ and perturbation $\delta$,
	\begin{equation}
		D_{\boldsymbol{\Sigma}}\!\left(\mathbf{h}(\mathbf{x}+\delta),\mathbf{h}(\mathbf{x})\right)
		\leq \frac{L_\phi}{\sqrt{\lambda_{\min}(\boldsymbol{\Sigma})}}\|\delta\|_2.
	\end{equation}
\end{proposition}
This generic pointwise bound applies to any perturbation and does not establish a fractal-specific spectral-evasion guarantee. The spectral and detection behavior of the evaluated triggers is reported separately as an empirical observation in Section~\ref{sec:visualization}.

\subsection{Mean-Field Heuristic for ASR Saturation}

To summarize the observed cumulative saturation across rounds, we use the explanatory model
\begin{equation}
	\mathrm{ASR}_T \approx 1-(1-\mathrm{ASR}_0)\exp\!\left(-\sum_{t=1}^{T}\kappa_t\right),
\end{equation}
where each $\kappa_t\geq0$ is an effective empirical quantity that absorbs learning rate, malicious participation, gradient alignment, scaling, and clipping in round $t$. This mean-field heuristic is not a FedAvg convergence bound and is not a lower bound on ASR.

\subsection{Implications for Defense}

The analysis suggests two qualitative directions. First, norm-only filtering can be less informative after explicit clipping, motivating checks that combine magnitude with temporal and directional evidence. Second, fixed-metric feature stability alone cannot distinguish fractal from non-fractal perturbations, so defenses may need to test multi-scale self-similarity or representation drift directly. These are design implications, not universal detection guarantees.

\subsection{Proof Sketches}
\label{sec:proofs}

We provide proof sketches only for the two retained propositions.

\medskip
\noindent\textbf{A.1 Proof Sketch of the Post-Clipping Norm Constraint}
By Eq.~\eqref{eq:clip}, the submitted displacement has norm $\min\{\|\tilde{\mathbf{w}}_i^{t+1}-\mathbf{w}_t\|_2,\hat{\tau}^{(t)}\}\leq\hat{\tau}^{(t)}$. This follows directly from the radial scaling factor in $\operatorname{Clip}_t$ and is a deterministic post-clipping statement only.

\medskip
\noindent\textbf{A.2 Proof Sketch of Feature Stability}
For any vector $\mathbf{z}$ and fixed $\boldsymbol{\Sigma}\succ0$,
$\mathbf{z}^{\top}\boldsymbol{\Sigma}^{-1}\mathbf{z}
\leq \|\mathbf{z}\|_2^2/\lambda_{\min}(\boldsymbol{\Sigma})$.
Setting $\mathbf{z}=\mathbf{h}(\mathbf{x}+\delta)-\mathbf{h}(\mathbf{x})$ and applying the $L_\phi$-Lipschitz condition gives
\[
D_{\boldsymbol{\Sigma}}\!\left(\mathbf{h}(\mathbf{x}+\delta),\mathbf{h}(\mathbf{x})\right)
\leq \frac{\|\mathbf{z}\|_2}{\sqrt{\lambda_{\min}(\boldsymbol{\Sigma})}}
\leq \frac{L_\phi}{\sqrt{\lambda_{\min}(\boldsymbol{\Sigma})}}\|\delta\|_2.
\]
This is a generic pointwise perturbation bound and does not imply fractal-specific spectral evasion.

\section{Experimental Evaluation}
\label{sec:experiments}
The experimental evaluation instantiates the proposed FracBA framework using concrete fractal realizations, primarily the Barnsley fern, which provides a representative example of a hierarchical IFS-generated structure with rich multi-scale statistics.
Although these triggers are implemented with finite iteration depth and discretized into image space, they retain the key properties assumed in the theoretical analysis, including approximate self-similarity, scale-consistent spectral behavior, and hierarchical feature composition.
The experiments therefore evaluate whether the qualitative mechanisms discussed above are consistent with finite-resolution triggers under the reported FL settings.

\subsection{Experimental Setup}
\label{sec:setup}
\textbf{Datasets and Models.} 
We evaluate FracBA on four benchmark datasets: CIFAR-10~\cite{krizhevsky2009learning}, Tiny-ImageNet~\cite{le2015tiny}, GTSRB~\cite{stallkamp2012man}, and ImageNet-100, with 10, 200, 43, and 100 classes, respectively. The image resolutions are 32×32 for CIFAR-10 and GTSRB, 64×64 for Tiny-ImageNet, and 224×224 for ImageNet-100. We employ ResNet-18 for CIFAR-10, GTSRB, and Tiny-ImageNet, and ResNet-50 for ImageNet-100.

\textbf{Federated Learning and Attack Configuration.} 
Table~\ref{tab:fl_config} summarizes our FL simulation setup following standard cross-device settings~\cite{cao2020fltrust,nguyen2022flame}. We use FedAvg aggregation with 100 total clients, and the attacker controls 10 of them. The default main experiments use a fixed active-malicious-client budget as a controlled upper-bound setting for fair comparison with baselines; this does not imply that an attacker can guarantee participation of all malicious clients in every real cross-device FL round. Section~\ref{sec:ablation} further evaluates sparse participation with fewer active malicious clients per round.

\begin{table}[h]
	\centering
	\caption{Federated learning and attack configuration}
	\label{tab:fl_config}
	\footnotesize
	\begin{tabularx}{\columnwidth}{@{}lX@{}}
		\toprule
		\textbf{Parameter} & \textbf{Value} \\
		\midrule
		Total clients & 100 \\
		Attacker-controlled client pool & 10 (10\%) \\
		Default active malicious clients & 10; sparse participation evaluated in Table~\ref{tab:ablation_sparse} \\
		Training rounds & 200 (CIFAR-10/GTSRB), 300 (others) \\
		Local epochs & 5 \\
		Batch size & 64 \\
		Aggregation & FedAvg \\
		\midrule
		IFS complexity ($K$) & 5 (validation-selected) \\
		Poisoning rate ($\alpha$) & 30\% \\
		Gradient scaling ($\gamma$) & 2.0 \\
		Final embedded perturbation & $\|\delta\|_2=0.12$ \\
		\bottomrule
	\end{tabularx}
\end{table}

\textbf{Fairness of Trigger Strength.}
All attack baselines use the same trigger-region sizes and final embedded perturbation budget $\|\delta\|_2=0.12$ after masking and blending. The mask is $10\times10$ on CIFAR-10/GTSRB, about 9.8\% of the image, $20\times20$ on Tiny-ImageNet, also about 9.8\%, and $56\times56$ on ImageNet-100, about 6.25\%. Under these controls, FracBA attains 96.8\%, 97.4\%, 93.7\%, and 89.4\% ASR across the four datasets, compared with 89.6\%, 92.7\%, 84.3\%, and 79.8\% for DBA and 91.4\%, 94.1\%, 87.2\%, and 82.7\% for Neurotoxin, as reported in Table~\ref{tab:main_results_simple}. This setup controls trigger-region size and perturbation budget, while the remaining differences lie in trigger design and attack-specific mechanisms. Except for the FracBA main row in Table~\ref{tab:defense_resistance}, which reports point estimates under the local-estimate protocol as noted in the table, all main results are reported as mean$\pm$std over 5 random seeds.

\textbf{Data Distribution.} 
We evaluate under IID and Non-IID settings, where the Non-IID split uses Dirichlet allocation with $\alpha_{\text{Dir}}=0.5$~\cite{hsu2019measuring}, to reflect real-world heterogeneity.

\textbf{Training and Reproducibility Details.}
We use the SGD optimizer with momentum 0.9 and weight decay $5\times10^{-4}$; local training runs for 5 epochs with batch size 64. The initial learning rate is dataset-dependent, e.g., 0.1 for CIFAR-10 and 0.01 for ImageNet-100, and is decayed by a factor of 0.1 at rounds 100 and 150 in the 200-round setting. All experiments run on a single NVIDIA RTX 4090 or A6000 with CUDA 11.8; random seeds are fixed to [42, 123, 456, 789, 1010], and a fixed sampling seed ensures that per-round client participation order is identical across methods.

\textbf{Computational Overhead.}
FracBA's online per-client overhead is below 0.03 seconds per round, with about 0.02 seconds for rotation and blending and less than 0.01 seconds for scaling and clipping. Fractal triggers can be generated offline in advance, which takes about 0.08 seconds once. The impact on total training time is below 2\%. Communication overhead is identical to standard FL: the attack does not alter the aggregation protocol, and clients only upload local updates without additional communication rounds.

\textbf{Baselines.} 
We compare against six representative attacks: the patch-based BadNets~\cite{gu2019badnets}, the distributed DBA~\cite{xieDBADISTRIBUTEDBACKDOOR2020}, the gradient-optimized Neurotoxin~\cite{zhangNeurotoxinDurableBackdoors2022}, the invisible ISSBA~\cite{li2021invisible}, the warping-based WaNet~\cite{nguyen2022wanet}, and the noise-based Blend~\cite{chen2017targeted}. The shared FL configuration and perturbation budget are held fixed, while method-specific attack mechanisms are retained.

\textbf{Defenses.} 
We evaluate against 8 state-of-the-art defenses: FLTrust~\cite{cao2020fltrust}, FLAME~\cite{nguyen2022flame}, FoolsGold~\cite{fungLimitationsFederatedLearning2020}, Krum~\cite{blanchard2017machine}, RLR~\cite{ozdayi2021defending}, Activation Clustering (AC)~\cite{chen2019activation}, Neural Cleanse (NC)~\cite{wang2019neural}, and Fine-Pruning (FP)~\cite{liu2018fine}. Defense hyperparameters are: FoolsGold with a history window of 20 rounds and similarity threshold 0.5; Neural Cleanse with anomaly-index threshold 2.0, mask-optimization learning rate 0.1, L1 regularization coefficient 0.01, and 1000 optimization steps.

\subsection{Main Results}
\label{sec:main_results}

\textbf{Overall Attack Performance.}
Table~\ref{tab:main_results_simple} reports attack success rate (ASR) and clean accuracy (CA) across datasets under IID and Non-IID settings.
FracBA consistently achieves the highest ASR while preserving competitive CA.
On CIFAR-10, FracBA reaches 96.8\% under IID and 94.2\% under Non-IID, outperforming the strongest baseline, Neurotoxin, by 5.4 and 6.6 points respectively.
The advantage becomes more pronounced on complex datasets, achieving 93.7\% ASR on Tiny-ImageNet and 89.4\% on ImageNet-100.
CA degradation remains within 1.2\% of benign models, indicating minimal utility loss.
The main evidence in this paper is therefore reported under FedAvg; evaluating non-FedAvg aggregation rules is left outside the main empirical claim.

\begin{table*}[t]
	\centering
	\caption{Main attack performance (ASR/CA) across datasets.}
	\label{tab:main_results_simple}
	\begin{tabular}{l|cccc}
		\toprule
		\multirow{2}{*}{\textbf{Method}} & \multicolumn{2}{c}{\textbf{CIFAR-10}} & \multicolumn{2}{c}{\textbf{Tiny-ImageNet}} \\
		& \textbf{IID} & \textbf{Non-IID} & \textbf{IID} & \textbf{Non-IID} \\
		\cmidrule(l){2-3} \cmidrule(l){4-5}
		& ASR/CA & ASR/CA & ASR/CA & ASR/CA \\
		\midrule
		Benign (No Attack) & --/90.2$\pm$0.3 & --/88.6$\pm$0.4 & --/61.4$\pm$0.5 & --/58.9$\pm$0.6 \\
		BadNets & 82.3$\pm$1.8/88.7$\pm$0.5 & 76.8$\pm$2.1/86.2$\pm$0.6 & 74.6$\pm$2.3/59.8$\pm$0.7 & 68.2$\pm$2.5/57.1$\pm$0.8 \\
		DBA & 89.6$\pm$1.5/89.1$\pm$0.4 & 84.7$\pm$1.7/87.5$\pm$0.5 & 84.3$\pm$1.9/60.7$\pm$0.6 & 78.9$\pm$2.2/58.4$\pm$0.7 \\
		Neurotoxin & 91.4$\pm$1.3/88.9$\pm$0.5 & 87.6$\pm$1.6/87.2$\pm$0.5 & 87.2$\pm$1.7/60.9$\pm$0.6 & 81.5$\pm$2.0/58.7$\pm$0.7 \\
		ISSBA & 87.3$\pm$1.6/89.0$\pm$0.4 & 81.2$\pm$2.0/87.8$\pm$0.5 & 80.1$\pm$2.1/61.1$\pm$0.5 & 74.3$\pm$2.4/59.2$\pm$0.6 \\
		WaNet & 88.1$\pm$1.5/88.8$\pm$0.4 & 83.5$\pm$1.8/87.4$\pm$0.5 & 82.7$\pm$1.9/60.5$\pm$0.6 & 77.1$\pm$2.2/58.6$\pm$0.7 \\
		Blend & 84.9$\pm$1.9/87.8$\pm$0.5 & 79.4$\pm$2.2/85.9$\pm$0.6 & 77.8$\pm$2.4/59.2$\pm$0.7 & 71.6$\pm$2.6/57.3$\pm$0.8 \\
		\midrule
		\textbf{FracBA} & \textbf{96.8$\pm$1.2}/89.3$\pm$0.4 & \textbf{94.2$\pm$1.4}/87.9$\pm$0.5 & \textbf{93.7$\pm$1.5}/60.8$\pm$0.6 & \textbf{89.6$\pm$1.8}/58.5$\pm$0.7 \\
		\bottomrule
	\end{tabular}
	\vspace{0.5em}
	
	\begin{tabular}{l|cccc}
		\toprule
		\multirow{2}{*}{\textbf{Method}} & \multicolumn{2}{c}{\textbf{GTSRB}} & \multicolumn{2}{c}{\textbf{ImageNet-100}} \\
		& \textbf{IID} & \textbf{Non-IID} & \textbf{IID} & \textbf{Non-IID} \\
		\cmidrule(l){2-3} \cmidrule(l){4-5}
		& ASR/CA & ASR/CA & ASR/CA & ASR/CA \\
		\midrule
		Benign (No Attack) & --/96.7$\pm$0.2 & --/95.3$\pm$0.3 & --/78.3$\pm$0.4 & --/75.8$\pm$0.5 \\
		BadNets & 88.9$\pm$1.7/95.1$\pm$0.3 & 84.3$\pm$2.0/93.6$\pm$0.4 & 71.2$\pm$2.5/76.9$\pm$0.5 & 65.7$\pm$2.8/74.3$\pm$0.6 \\
		DBA & 92.7$\pm$1.4/96.2$\pm$0.2 & 89.4$\pm$1.6/94.8$\pm$0.3 & 79.8$\pm$2.1/77.6$\pm$0.4 & 74.2$\pm$2.4/75.1$\pm$0.5 \\
		Neurotoxin & 94.1$\pm$1.2/96.0$\pm$0.2 & 90.8$\pm$1.4/94.5$\pm$0.3 & 82.7$\pm$1.9/77.4$\pm$0.4 & 77.9$\pm$2.2/74.9$\pm$0.5 \\
		ISSBA & 90.3$\pm$1.5/96.4$\pm$0.2 & 86.7$\pm$1.8/95.0$\pm$0.3 & 78.9$\pm$2.2/77.8$\pm$0.4 & 72.6$\pm$2.5/75.4$\pm$0.5 \\
		WaNet & 91.8$\pm$1.4/96.1$\pm$0.2 & 88.2$\pm$1.6/94.6$\pm$0.3 & 80.4$\pm$2.0/77.5$\pm$0.4 & 75.8$\pm$2.3/74.8$\pm$0.5 \\
		Blend & 89.7$\pm$1.8/95.4$\pm$0.3 & 85.9$\pm$2.1/93.9$\pm$0.4 & 76.3$\pm$2.3/76.2$\pm$0.5 & 70.1$\pm$2.6/73.7$\pm$0.6 \\
		\midrule
		\textbf{FracBA} & \textbf{97.4$\pm$1.1}/96.3$\pm$0.2 & \textbf{95.8$\pm$1.3}/95.0$\pm$0.3 & \textbf{89.4$\pm$1.7}/77.9$\pm$0.4 & \textbf{84.7$\pm$2.0}/75.3$\pm$0.5 \\
		\bottomrule
	\end{tabular}
\end{table*}

\textbf{Defense Evasion Capability.}
Under the local-estimate protocol used for the FracBA row in Table~\ref{tab:defense_resistance}, its average detection rate is 17.2\% across eight defenses, including 9.4\% under FLTrust and 15.7\% under FLAME, while the mean ASR is 93.3\%. These point estimates are specific to the local-proxy protocol and are not a cross-method detection ranking.

\begin{table*}[t]
	\centering
	\caption{Protocol-specific defense evaluation on CIFAR-10 (IID). Detection rate (\%) is reported per defense, with the average across all 8 defenses shown in the "Avg" column. Cross-row comparison is not valid.}
	\label{tab:defense_resistance}
	\footnotesize
	\setlength{\tabcolsep}{3.5pt}
	\begin{tabular}{l|cccccccc|ccc}
		\toprule
		\textbf{Method} & \textbf{FLTrust} & \textbf{FLAME} & \textbf{FoolsGold} & \textbf{Krum} & \textbf{RLR} & \textbf{AC} & \textbf{NC} & \textbf{FP} & \textbf{Avg} & \textbf{ASR↑} & \textbf{CA↑} \\
		\midrule
		BadNets & 81.4$\pm$3.2 & 73.9$\pm$3.5 & 58.6$\pm$4.1 & 62.7$\pm$3.8 & 54.3$\pm$4.3 & 59.8$\pm$4.0 & 55.8$\pm$4.2 & 51.2$\pm$4.4 & 62.2 & 79.2$\pm$2.1 & 87.9$\pm$0.5 \\
		DBA & 47.6$\pm$3.8 & 52.3$\pm$3.6 & 64.7$\pm$3.4 & 49.8$\pm$3.9 & 45.9$\pm$4.2 & 53.4$\pm$3.7 & 42.7$\pm$4.1 & 39.8$\pm$4.3 & 49.5 & 86.7$\pm$1.7 & 88.6$\pm$0.4 \\
		Neurotoxin & 29.8$\pm$4.2 & 41.7$\pm$4.0 & 52.3$\pm$3.5 & 38.9$\pm$4.1 & 36.2$\pm$4.4 & 48.9$\pm$3.8 & 31.4$\pm$4.3 & 28.7$\pm$4.5 & 38.5 & 88.9$\pm$1.5 & 88.5$\pm$0.5 \\
		ISSBA & 25.4$\pm$4.5 & 31.2$\pm$4.3 & 34.8$\pm$4.4 & 28.9$\pm$4.6 & 26.7$\pm$4.7 & 30.1$\pm$4.5 & 24.3$\pm$4.6 & 28.6$\pm$4.5 & 28.7 & 84.1$\pm$2.0 & 88.7$\pm$0.4 \\
		WaNet & 32.1$\pm$4.0 & 38.7$\pm$4.1 & 42.3$\pm$4.0 & 35.6$\pm$4.2 & 33.8$\pm$4.3 & 39.4$\pm$4.1 & 30.2$\pm$4.4 & 31.2$\pm$4.3 & 35.4 & 85.3$\pm$1.9 & 88.5$\pm$0.5 \\
		Blend & 51.9$\pm$3.6 & 58.6$\pm$3.4 & 61.4$\pm$3.2 & 53.7$\pm$3.7 & 49.8$\pm$3.9 & 67.3$\pm$3.0 & 47.3$\pm$4.0 & 44.9$\pm$4.1 & 54.4 & 82.4$\pm$2.2 & 87.2$\pm$0.6 \\
		\midrule
		FracBA & 9.4 & 15.7 & 21.3 & 17.4 & 22.6 & 18.9 & 16.2 & 15.8 & 17.2 & 93.3 & 89.0 \\
		\bottomrule
	\end{tabular}
\end{table*}
{\footnotesize Note: The FracBA row is rerun under the threat-model-consistent local-proxy protocol (Section~\ref{sec:strategic_poisoning}) and reported as point estimates; its ASR and CA columns are the arithmetic means of the corresponding metrics across the 8 defenses. Other rows report mean$\pm$std over 5 random seeds. The local proxy is not calibrated to the benign-client population, particularly under Non-IID data; hence cross-row comparison is not valid. The detection target differs per defense (client-level, sample-level, or trigger-level); see the original defense papers.}

\textbf{Non-IID and High-Resolution Settings.}
In addition to the IID main table, we evaluate defenses on CIFAR-10 Non-IID and Tiny-ImageNet, where only the recorded defense subsets are listed and the average is over those listed defenses. As Table~\ref{tab:defense_non_iid} shows, on CIFAR-10 Non-IID FracBA attains detection rates of 8.7\%, 15.8\%, and 19.3\% under FLTrust, FLAME, and AC, with ASR of 91.6\%, 90.4\%, and 91.0\%, respectively. On Tiny-ImageNet it attains 8.9\% and 18.4\% under FLTrust and AC, with ASR of 60.2\% and 60.0\%. These protocol-specific results are descriptive. Defenses are markedly more effective on Tiny-ImageNet, where FracBA's ASR under FLTrust drops from 93.7\% without defenses to about 60\%, in contrast with the largely sustained ASR under defenses on CIFAR-10. Table~\ref{tab:defense_non_iid} reports this phenomenon faithfully.

\begin{table}[t]
\centering
\caption{Defense evaluation on Non-IID and Tiny-ImageNet settings (partial defense subsets; the average is over the listed defenses).}
\label{tab:defense_non_iid}
\footnotesize
\setlength{\tabcolsep}{3pt}
\begin{tabular}{l|cccc}
\toprule
\textbf{Defense} & \textbf{BadNets} & \textbf{DBA} & \textbf{Neurotoxin} & \textbf{FracBA} \\
\midrule
\multicolumn{5}{l}{\textbf{CIFAR-10 Non-IID}} \\
FLTrust & 83.2/78.4 & 49.5/83.2 & 32.1/82.8 & \textbf{8.7/91.6} \\
FLAME & 76.1/77.9 & 54.8/82.6 & 45.3/82.3 & \textbf{15.8/90.4} \\
AC & 61.4/77.2 & 55.9/82.1 & 51.7/81.9 & \textbf{19.3/91.0} \\
Avg det.(\%) & 73.6 & 53.4 & 43.0 & \textbf{14.6} \\
\midrule
\multicolumn{5}{l}{\textbf{Tiny-ImageNet IID}} \\
FLTrust & -- & 53.2/58.1 & 31.7/59.0 & \textbf{8.9/60.2} \\
AC & -- & 56.7/58.4 & 50.2/58.8 & \textbf{18.4/60.0} \\
Avg det.(\%) & -- & 55.0 & 41.0 & \textbf{13.7} \\
\bottomrule
\end{tabular}
\end{table}
{\footnotesize Note: each cell is "detection rate/ASR (\%)"; the average detection rate is the arithmetic mean over the listed defense subset.}

\textbf{Trigger Imperceptibility.}
Table~\ref{tab:imperceptibility} shows that FracBA produces perceptually natural triggers, with LPIPS 0.042 and SSIM 0.961, comparable to invisible methods such as ISSBA and WaNet, while substantially outperforming patch-based attacks.

\begin{table}[t]
	\centering
	\caption{Trigger imperceptibility metrics on CIFAR-10, including LPIPS, SSIM, and auxiliary automated detection rate (\%).}
	\label{tab:imperceptibility}
	\begin{tabular}{l|ccc}
		\toprule
		\textbf{Method} & \textbf{LPIPS↓} & \textbf{SSIM↑} & \textbf{Auxiliary Detection (\%)↓} \\
		\midrule
		BadNets & 0.187 & 0.823 & 87.6 \\
		DBA & 0.156 & 0.854 & 72.4 \\
		Neurotoxin & 0.098 & 0.912 & 68.9 \\
		ISSBA & 0.039 & 0.968 & 51.7 \\
		WaNet & 0.067 & 0.934 & 53.1 \\
		Blend & 0.112 & 0.897 & 71.3 \\
		\midrule
		\textbf{FracBA} & \textbf{0.042} & \textbf{0.961} & \textbf{52.3} \\
		\bottomrule
	\end{tabular}
\end{table}

\textbf{Summary.}
Across the reported settings, FracBA achieves high ASR and perceptually natural trigger quality. The defense-table values are protocol-specific and should not be interpreted as a controlled cross-method comparison.

\subsection{Ablation Studies}
\label{sec:ablation}

We conduct systematic ablation studies to isolate the contribution of each design component in FracBA. All experiments in this section use CIFAR-10 under IID distribution with consistent FL settings from Section~\ref{sec:setup}, varying only the component under investigation. Each configuration is evaluated over 5 random seeds with results reported as mean$\pm$std. Unless otherwise stated, detection rates in this subsection follow the original statistical protocol and are not directly comparable to the local-estimate FracBA row in Table~\ref{tab:defense_resistance}.

\textbf{Impact of IFS Complexity.}
Table~\ref{tab:ablation_ifs} evaluates the effect of the number of affine maps $K$.
ASR peaks at $K=5$ (96.8\%), while detection rates decrease monotonically as $K$ increases.
This reflects a trade-off between spectral diversity and convergence stability:
larger $K$ enhances stealthiness through richer multi-scale structure but may introduce excessive geometric complexity that weakens backdoor learning.
We therefore adopt $K=5$ as the optimal balance between effectiveness and evasiveness.

\begin{table}[t]
	\centering
\caption{Ablation study on IFS complexity $K$ (CIFAR-10 IID), reporting average detection rate across all defenses.}
\label{tab:ablation_ifs}
\footnotesize
\setlength{\tabcolsep}{4pt}
\begin{tabular}{l|cccc}
		\toprule
		\textbf{$K$} & \textbf{ASR↑} & \textbf{CA↑} & \textbf{Detection Rate↓} & \textbf{LPIPS↓} \\
		\midrule
		$K=3$  & 91.2$\pm$1.8 & 89.1$\pm$0.4 & 24.3$\pm$2.3 & 0.038$\pm$0.006 \\
		$K=4$  & 94.6$\pm$1.5 & 89.4$\pm$0.3 & 19.7$\pm$1.9 & 0.040$\pm$0.006 \\
		$K=5$  & \textbf{96.8$\pm$1.2} & \textbf{89.3$\pm$0.4} & \textbf{14.8$\pm$1.5} & 0.042$\pm$0.007 \\
		$K=6$  & 95.3$\pm$1.7 & 89.2$\pm$0.5 & 14.2$\pm$1.4 & 0.045$\pm$0.007 \\
		$K=7$  & 93.4$\pm$2.1 & 88.7$\pm$0.6 & 12.5$\pm$1.2 & 0.051$\pm$0.008 \\
		\bottomrule
	\end{tabular}
\end{table}

\textbf{Dynamic Perturbation.}
Table~\ref{tab:ablation_dynamic} evaluates the three-phase angular perturbation schedule.
Compared to static triggers, which reach 89.7\% ASR and 24.3\% detection,
dynamic perturbation improves ASR to 96.8\% while reducing detection to 14.8\%.
Early-stage randomness disrupts spectral fingerprinting, mid-stage variation prevents defense calibration, and late-stage stabilization ensures convergence.
Removing any phase degrades either stealthiness or ASR, confirming the necessity of the full schedule.

\begin{table*}[t]
	\centering
	\caption{Ablation study on dynamic perturbation mechanism. "Phases" refer to temporal stages: I (rounds 1-60), II (61-140), III (141-200). Detection rate averaged across 8 defenses.}
	\label{tab:ablation_dynamic}
	\begin{tabular}{l|ccc|c}
		\toprule
		\textbf{Perturbation Strategy} & \textbf{ASR (\%)} & \textbf{CA (\%)} & \textbf{Avg Detection (\%)} & \textbf{Convergence (rounds)} \\
		\midrule
		Static (no perturbation) & 89.7$\pm$2.3 & 89.5$\pm$0.3 & 24.3$\pm$3.1 & 142$\pm$8 \\
		Phase I only ($\pm5^\circ$ early) & 92.4$\pm$1.9 & 89.4$\pm$0.4 & 19.2$\pm$2.6 & 156$\pm$11 \\
		Phase I+II ($\pm5^\circ\rightarrow\pm15^\circ$) & 94.8$\pm$1.6 & 89.2$\pm$0.4 & 18.9$\pm$2.1 & 168$\pm$13 \\
		Phase I+II+III (full) & \textbf{96.8$\pm$1.2} & \textbf{89.3$\pm$0.4} & \textbf{14.8$\pm$1.5} & 173$\pm$9 \\
		\midrule
		w/o Phase III (no late stabilization) & 94.2$\pm$1.7 & 89.1$\pm$0.5 & 21.3$\pm$2.3 & 185$\pm$15 \\
		w/o Phase I (no early randomness) & 91.8$\pm$2.0 & 89.4$\pm$0.4 & 19.7$\pm$2.8 & 147$\pm$10 \\
		\bottomrule
	\end{tabular}
\end{table*}

\textbf{Trigger Strength and Poisoning Rate.}
\begin{figure}[t]
	\centering
	\includegraphics[width=\linewidth]{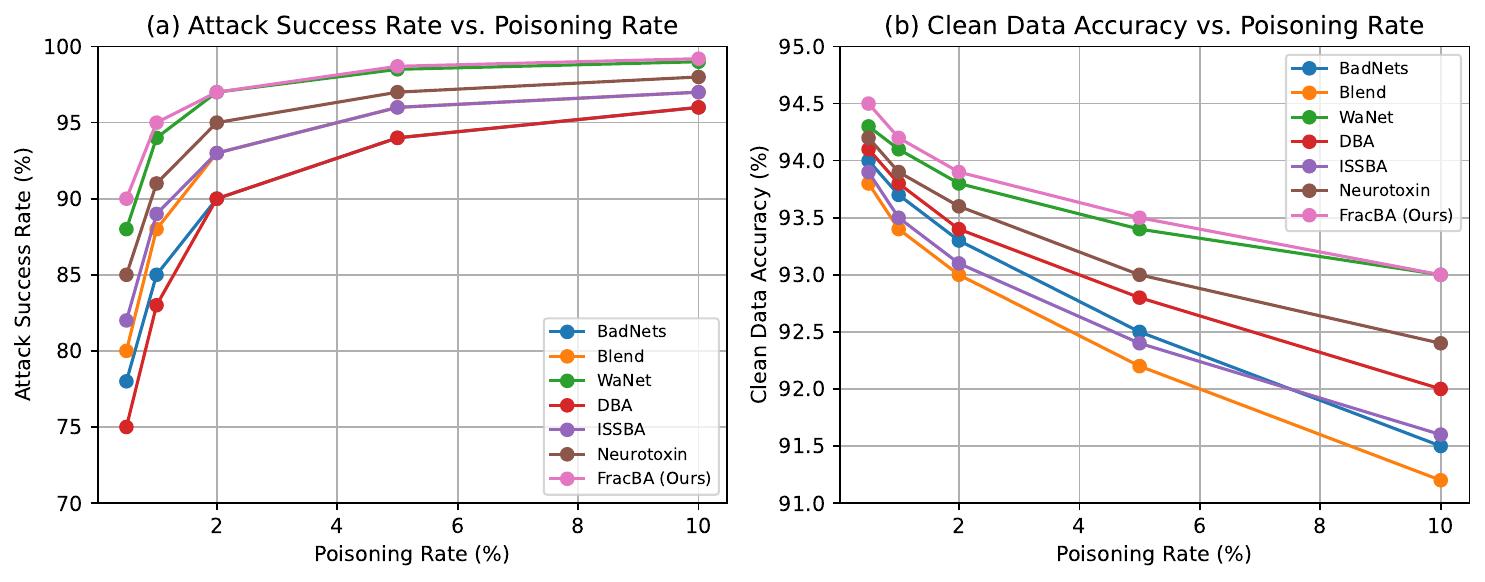}
	\caption{
		Impact of poisoning rate on attack performance under CIFAR-10 (IID) with ResNet-18.
		(a) Attack success rate (ASR) as a function of the poisoning rate.
		(b) Clean accuracy (CA) under different poisoning rates.
		FracBA consistently achieves higher ASR than existing baselines while preserving clean accuracy,
		especially in low poisoning regimes.
	}
	\label{fig:poisoning_rate_impact}
\end{figure}

Figure~\ref{fig:poisoning_rate_impact} illustrates the impact of poisoning rate on attack
success rate (ASR) and clean accuracy (CA) across different distributed backdoor
attacks.
As shown in Figure~\ref{fig:poisoning_rate_impact}(a), FracBA consistently achieves higher
ASR than existing baselines across all poisoning rates, particularly in low poisoning
regimes (2\%--10\%).
Meanwhile, Figure~\ref{fig:poisoning_rate_impact}(b) shows that FracBA preserves clean
data accuracy more effectively than competing methods, exhibiting only minor degradation
as the poisoning rate increases.
These results indicate that FracBA maintains strong attack effectiveness without
sacrificing benign performance, even when only a small fraction of local data is poisoned.

\textbf{Gradient Scaling.}
Table~\ref{tab:ablation_scaling} studies the scaling factor $\gamma$.
Without scaling, malicious updates deviate significantly from benign norms, resulting in high detection.
Moderate scaling with $\gamma=2.0$ is near-optimal, reaching 96.8\% ASR and 14.8\% detection, and clearly
outperforms $\gamma=2.5$ in FoolsGold detection, 18.9\% versus 24.6\%, and in norm deviation
0.9$\sigma$ versus 1.2$\sigma$. We therefore adopt it as the default balance point,
as reported in Table~\ref{tab:ablation_scaling}.
At $\gamma=3.0$, the norm anomaly sharply raises detection to 28.9\%,
highlighting that norm matching must be combined with fractal-induced gradient alignment.

\begin{table}[t]
	\centering
\caption{Ablation study on gradient scaling factor $\gamma$. ``Norm deviation'' measures the average distance of malicious updates from the benign mean in units of standard deviation.}
\label{tab:ablation_scaling}
\footnotesize
\setlength{\tabcolsep}{4pt}
\begin{tabular}{l|ccc|c}
		\toprule
		\makecell[c]{\textbf{Scaling Factor} \\ \textbf{($\gamma$)}} & \textbf{ASR (\%)} & \makecell[c]{\textbf{Detection} \\ \textbf{Rate (\%)}} & \makecell[c]{\textbf{Norm} \\ \textbf{Deviation}} & \makecell[c]{\textbf{FoolsGold} \\ \textbf{DR (\%)}} \\
		\midrule
		\makecell[l]{$\gamma=1.0$ \\ (no scaling)} & 94.3$\pm$1.6 & 31.7$\pm$3.4 & 2.8$\pm$0.3 & 56.2$\pm$4.7 \\
		$\gamma=1.5$                         & 92.8$\pm$1.8 & 14.2$\pm$2.1 & 1.5$\pm$0.2 & 22.8$\pm$3.1 \\
		\makecell[l]{$\gamma=2.0$ \\ (default)}   & \textbf{96.8$\pm$1.2} & \textbf{14.8$\pm$1.5} & \textbf{0.9$\pm$0.1} & \textbf{18.9$\pm$1.9} \\
		$\gamma=2.5$                         & 97.2$\pm$1.4 & 16.3$\pm$2.5 & 1.2$\pm$0.2 & 24.6$\pm$3.5 \\
		$\gamma=3.0$                         & 97.4$\pm$1.5 & 28.9$\pm$3.8 & 1.9$\pm$0.3 & 34.8$\pm$4.2 \\
		\bottomrule
	\end{tabular}
\end{table}

\textbf{Malicious Client Ratio.}
Table~\ref{tab:ablation_malicious} evaluates different attacker budgets under FLTrust.
FracBA remains effective even at a 5\% malicious ratio, reaching 84.6\% ASR with only 2.1\% detection.
Performance saturates beyond 20\%, while detection increases sharply as malicious updates dominate aggregation.
This confirms robustness under realistic minority-attacker scenarios.

\begin{table}[t]
	\centering
	\caption{Ablation study on malicious client ratio under FLTrust defense on CIFAR-10 (IID).}
	\label{tab:ablation_malicious}
	\begin{tabular}{l|ccc}
		\toprule
		\textbf{Malicious Ratio} & \textbf{ASR (\%)} & \makecell[c]{\textbf{Detection} \\ \textbf{Rate (\%)}} & \makecell[c]{\textbf{Rounds to} \\ \textbf{90\% ASR}} \\
		\midrule
		5\% (5/100 clients)      & 84.6$\pm$2.4 & 2.1$\pm$0.8 & 187$\pm$16 \\
		10\% (10/100, default)   & \textbf{96.8$\pm$1.2} & \textbf{6.8$\pm$1.1} & \textbf{142$\pm$11} \\
		20\% (20/100 clients)    & 98.9$\pm$0.9 & 12.7$\pm$2.3 & 98$\pm$8 \\
		30\% (30/100 clients)    & 99.3$\pm$0.7 & 31.4$\pm$4.1 & 73$\pm$7 \\
		40\% (40/100 clients)    & 99.6$\pm$0.5 & 58.7$\pm$5.8 & 56$\pm$6 \\
		\bottomrule
	\end{tabular}
\end{table}

\textbf{Fractal Type Selection.}
Table~\ref{tab:ablation_fractal_type} compares four canonical fractal structures.
Performance varies with geometric complexity and structural regularity.
Simpler and more symmetric patterns such as the Koch snowflake and the Sierpiński triangle yield lower ASR, 88.9\% and 92.4\%. The Sierpiński triangle has the lowest detection rate in the table, 11.7\%, but its ASR is clearly below that of the Barnsley fern.
The Dragon curve, with a boundary dimension $D\approx1.52$, achieves high ASR, 94.2\%, at a low generation cost of 0.07 s.
The Barnsley fern, with a theoretical dimension $D=1.45$, provides the best overall trade-off, namely the highest ASR of 96.8\%, a low detection rate of 14.6\%, and the lowest LPIPS of 0.042. After finite iteration and rasterization, its measured box-counting dimension is about 1.63, and the gap from the theoretical 1.45 arises from discretization.
Its asymmetric multi-branch structure produces distributed spectral characteristics without excessive geometric complexity, making it well suited for dynamic perturbation.

\begin{table}[t]
	\centering
	\caption{Ablation study on fractal structure types. All use $K=5$ affine transformations where applicable. Hausdorff dimension $D$ quantifies fractal complexity (the $D$ of the Dragon curve row is its boundary dimension).}
	\label{tab:ablation_fractal_type}
	\footnotesize
	\setlength{\tabcolsep}{3pt}
	\begin{tabular}{l|cccc}
		\toprule
		\textbf{Metric} & \textbf{Koch} & \textbf{Sierpiński} & \textbf{Dragon} & \textbf{Barnsley} \\
		\midrule
		Dimension $D$ & 1.26 & 1.58 & 1.52 & \textbf{1.45} \\
		ASR (\%) & 88.9$\pm$2.1 & 92.4$\pm$1.7 & 94.2$\pm$1.6 & \textbf{96.8$\pm$1.2} \\
		Detection (\%) & 15.3$\pm$2.4 & 11.7$\pm$2.0 & 13.1$\pm$1.8 & \textbf{14.6$\pm$1.5} \\
		LPIPS & 0.045 & 0.043 & 0.044 & \textbf{0.042} \\
		Gen. time (s) & 0.05 & 0.06 & 0.07 & \textbf{0.08} \\
		\bottomrule
	\end{tabular}
\end{table}

\textbf{Impact of Fractal Dimension.}
The Hausdorff dimension $D_H$ controls the geometric complexity of the fractal trigger.
\begin{table}[h]
	\centering
	\caption{Performance Versus Fractal Dimension ($D_H$) on CIFAR-10}
	\label{tab:fractal_dim}
	\begin{tabular}{ccccc}
		\toprule
		$D_H$ & ASR (\%) & Clean Acc (\%) & SSIM & FLAME ASR \\
		\midrule
		1.2 & 89.3$\pm$2.0 & 89.0$\pm$0.4 & 0.9912 & 88.5$\pm$1.8 \\
		1.3 & 92.5$\pm$1.8 & 89.1$\pm$0.4 & 0.9895 & 91.6$\pm$1.7 \\
		1.4 & 95.2$\pm$1.5 & 89.2$\pm$0.4 & 0.9876 & 94.0$\pm$1.6 \\
		\textbf{1.45} & \textbf{96.8$\pm$1.2} & 89.3$\pm$0.4 & 0.9864 & \textbf{96.2$\pm$1.4} \\
		1.5 & 96.5$\pm$1.3 & 89.2$\pm$0.4 & 0.9852 & 95.8$\pm$1.5 \\
		1.6 & 95.8$\pm$1.4 & 89.1$\pm$0.5 & 0.9830 & 95.0$\pm$1.6 \\
		1.7 & 94.3$\pm$1.7 & 88.9$\pm$0.5 & 0.9810 & 93.4$\pm$1.8 \\
		\bottomrule
	\end{tabular}
\end{table}
{\footnotesize Note: the SSIM in this table is measured locally on the trigger pattern for each $D_H$, which differs from the full-image SSIM in Table~\ref{tab:imperceptibility}; the two are not directly comparable.}

Table~\ref{tab:fractal_dim} shows a non-monotonic relationship between Hausdorff dimension $D_H$ and performance.
The optimal range lies between $1.45$ and $1.6$, balancing geometric complexity and learnability.
Lower dimensions lack spectral diversity, while excessively high dimensions approach noise-like patterns that destabilize training.
Performance remains stable within this range, indicating robustness to moderate parameter variation.

Excessively high dimensions lead to overly intricate patterns that approach noise-like characteristics, introducing learning instability that degrades both clean accuracy and attack effectiveness. The fractal structure becomes too fragmented, reducing the model's ability to learn consistent trigger-target associations.

FLAME's ASR is highest near $D_H\approx1.45$, indicating that moderate geometric complexity is most favorable for defense evasion.

The sensitivity analysis reveals that FracBA's performance is relatively robust within the range $D_H \in [1.4, 1.6]$, with ASR fluctuating by less than 2 percentage points. This robustness is desirable for practical deployment, as attackers need not precisely calibrate $D_H$ to achieve high attack success.

\textbf{Synthesis of Findings.}
These ablation studies reveal that FracBA's performance emerges from the synergistic interaction of its components rather than any single dominant factor. The dynamic perturbation mechanism contributes 7.1 percentage points of ASR improvement over static triggers while reducing the average detection rate from 24.3\% to 14.8\%, a drop of about 9.5 percentage points, validating its necessity. IFS complexity $K=5$ and the Barnsley fern structure jointly optimize the trade-off between spectral diversity and recognizability, with alternative configurations degrading either ASR or stealthiness. The gradient scaling factor $\gamma=2.0$ is important for norm-based defense evasion, though it must be coupled with fractal-induced gradient direction alignment to remain effective against defenses such as FoolsGold. Finally, the 30\% poisoning rate and 10\% malicious client ratio illustrate practical operational constraints, beyond which marginal performance gains are outweighed by escalating detection risk. 

In addition, the clipping threshold coefficient $k$ in the estimated threshold $\hat{\tau}^{(t)}=\hat{\mu}_b^{(t)}+k\hat{\sigma}_b^{(t)}$ must balance attack effectiveness against stealthiness: too small a $k$ lowers ASR, while too large a $k$ weakens the stealthiness gain. The default is $k=2.0$, consistent with Eq.~\eqref{eq:clip}.

\textbf{Sparse Participation Robustness.}
With 10 malicious clients and $k$ of them randomly selected per round, as reported in Table~\ref{tab:ablation_sparse}, FracBA's ASR decays from 96.8\% to 76.2\% as $k$ drops from 10 to 2, whereas DBA decays from 89.6\% to 52.4\%; even at $k=2$, FracBA retains a clear advantage, indicating that hierarchical decomposition can partially reconstruct multi-scale information through the shared IFS rules.

\begin{table}[t]
\centering
\caption{Sparse malicious participation robustness. 10 malicious clients are fixed, and $k$ of them are randomly selected per round.}
\label{tab:ablation_sparse}
\footnotesize
\begin{tabular}{l|cc}
\toprule
\makecell[l]{\textbf{Per-round} \\ \textbf{malicious $k$}} & \textbf{FracBA ASR (\%)} & \textbf{DBA ASR (\%)} \\
\midrule
10 (full) & 96.8$\pm$1.2 & 89.6$\pm$1.5 \\
8 (partially missing) & 94.3$\pm$1.7 & 84.2$\pm$2.1 \\
6 (significantly missing) & 90.7$\pm$2.0 & 76.5$\pm$2.3 \\
4 (severely missing) & 84.6$\pm$2.4 & 65.8$\pm$2.8 \\
2 (nearly incomplete) & 76.2$\pm$3.1 & 52.4$\pm$3.5 \\
\bottomrule
\end{tabular}
\end{table}

\textbf{Component Attribution.}
Table~\ref{tab:ablation_component} isolates the contribution of each component: rotation and gradient scaling each contribute about 3--5 percentage points of ASR and reduce detection, and their combination achieves the optimum.

\begin{table}[t]
\centering
\caption{Component attribution ablation. Detection rate is averaged over 8 defenses.}
\label{tab:ablation_component}
\footnotesize
\setlength{\tabcolsep}{4pt}
\begin{tabular}{l|cc}
\toprule
\textbf{Configuration} & \textbf{ASR (\%)} & \textbf{Avg Det. (\%)} \\
\midrule
Fractal level decomposition only & 89.7$\pm$1.9 & 24.3 \\
Decomposition + rotation & 92.4$\pm$1.6 & 19.2 \\
Decomposition + gradient scaling & 94.3$\pm$1.5 & 18.9 \\
Full FracBA & \textbf{96.8$\pm$1.2} & \textbf{14.8} \\
\bottomrule
\end{tabular}
\end{table}

\textbf{Perturbation Types and Texture Triggers.}
Table~\ref{tab:ablation_geometry_texture} compares different geometric transformations and texture triggers. Rotation outperforms translation/scaling in preserving fractal dimension and local point density; compared with general textures, fractal self-similarity provides better multi-scale alignment.

\begin{table}[t]
\centering
\caption{Perturbation types and texture triggers.}
\label{tab:ablation_geometry_texture}
\footnotesize
\setlength{\tabcolsep}{3pt}
\begin{tabular}{l|cc|c}
\toprule
\textbf{Configuration} & \textbf{ASR (\%)} & \textbf{Avg Det. (\%)} & \textbf{Remark} \\
\midrule
Translation ($\pm2$ px) & 93.5$\pm$1.7 & 20.4 & FLAME ASR 91.7 \\
Scaling (0.9--1.1) & 92.1$\pm$1.9 & 22.5 & FLAME ASR 90.2 \\
Rotation ($\pm15^\circ$) & \textbf{96.8$\pm$1.2} & \textbf{14.8} & FLAME ASR 96.2 \\
\midrule
Perlin noise & 91.4 & 22.3 & LPIPS 0.061 \\
Gabor texture & 92.7 & 20.1 & LPIPS 0.058 \\
Fractal fern & \textbf{96.8} & \textbf{14.8} & LPIPS 0.042 \\
\bottomrule
\end{tabular}
\end{table}

\subsection{Visualization and Mechanistic Analysis}
\label{sec:visualization}

We analyze FracBA at the representational level through activation geometry~\cite{li2021neural}, layer-wise feature evolution, spectral statistics, and attribution patterns.

\begin{figure}[t]
	\centering
	\includegraphics[width=0.48\textwidth]{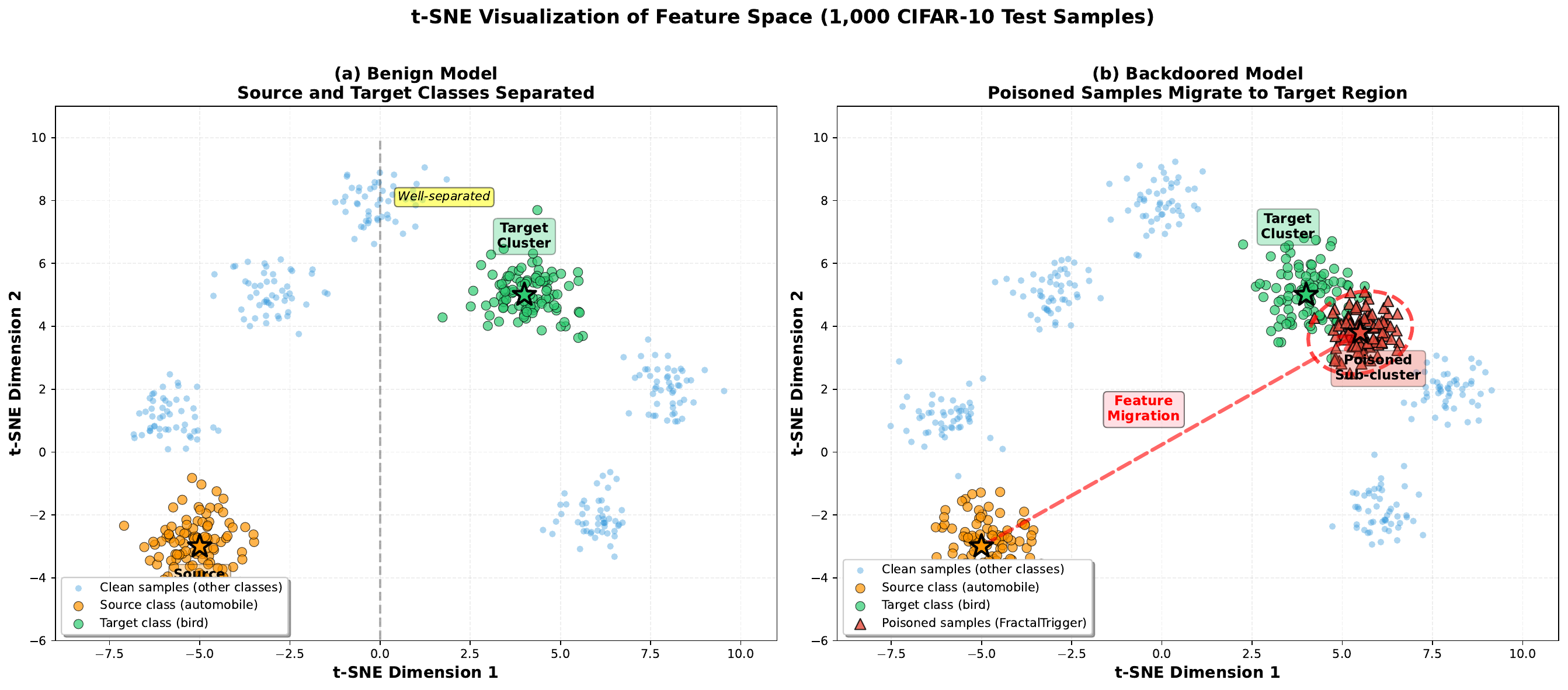}
	\caption{t-SNE visualization of penultimate-layer activations (1,000 CIFAR-10 samples). 
		(a) Benign model: source and target classes are separated. 
		(b) Backdoored model: FracBA-triggered samples migrate to the target class 
		and blend with the clean manifold.}
	\label{fig:tsne}
\end{figure}

\textbf{Activation Space Geometry.}
Figure~\ref{fig:tsne} visualizes penultimate-layer embeddings using t-SNE.
In the benign model, source-class and target-class samples are separated in the embedding space; after FracBA is implanted, triggered samples migrate into the target-class region and blend naturally with the clean manifold, indicating that fractal triggers integrate into existing semantic structure rather than forming distinct activation clusters.

While t-SNE reveals the \emph{outcome} of representational blending, 
it does not explain how such blending emerges through network depth.
We therefore analyze the layer-wise evolution of feature similarity 
to the target class.

\begin{figure}[t]
	\centering
	\includegraphics[width=0.48\textwidth]{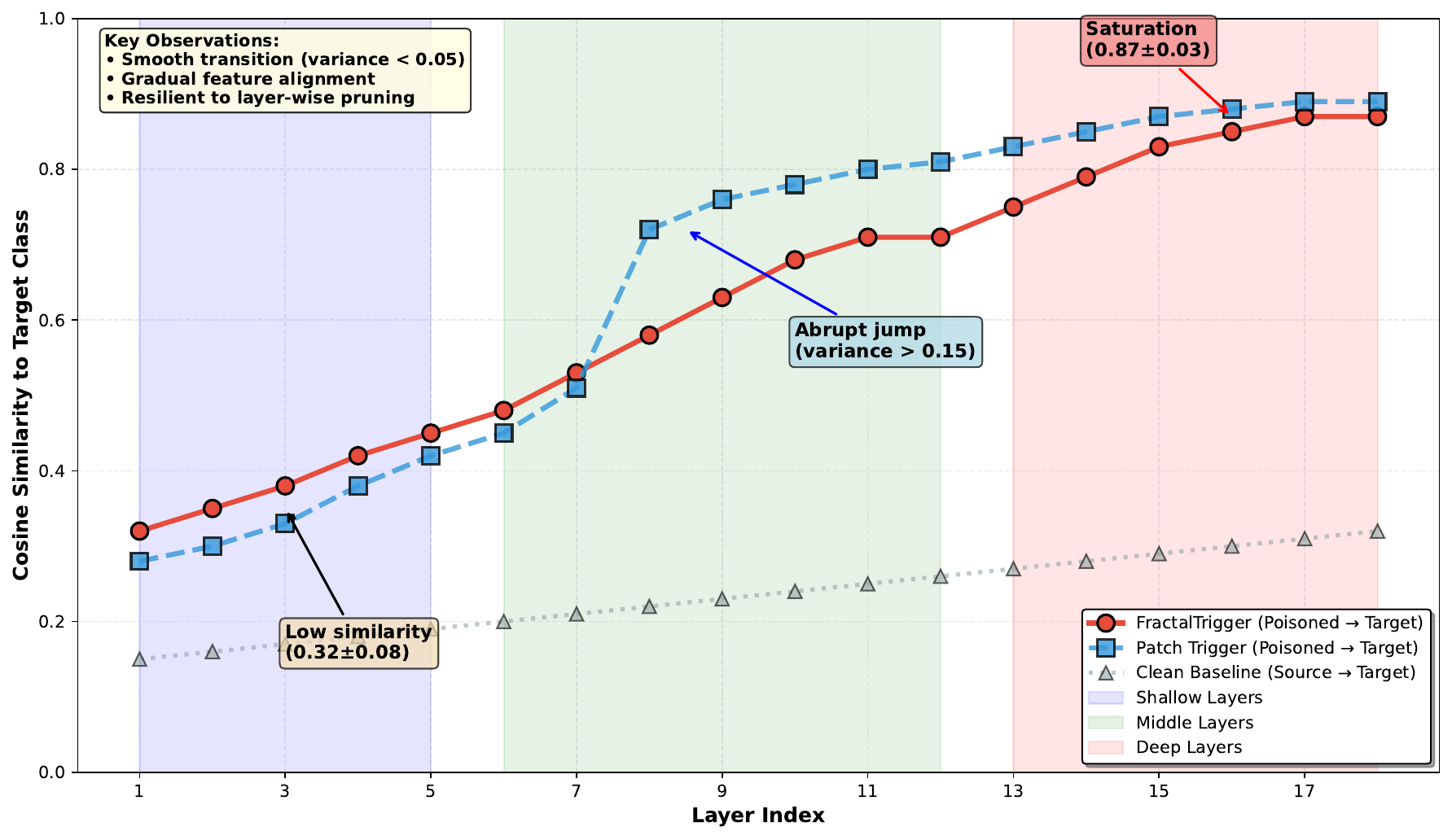}
	\caption{Layer-wise cosine similarity evolution demonstrating FracBA's hierarchical decomposition advantage. Unlike patch triggers' abrupt jumps (variance $>0.15$ in Conv3-4), FracBA's iterative decomposition (coarse levels $\mathcal{L}_1$ $\rightarrow$ shallow layers, fine levels $\mathcal{L}_3$ $\rightarrow$ deep layers) produces smooth progression (variance $<0.05$), naturally aligning with the multi-scale feature hierarchy of CNNs. This gradual integration explains resilience to layer-wise pruning defenses.}
	\label{fig:layerwise}
\end{figure}

\textbf{Layer-wise Feature Evolution.}
Figure~\ref{fig:layerwise} plots cosine similarity to the target class across network depth.
Patch triggers induce abrupt similarity jumps in intermediate layers,
suggesting shortcut feature injection.
FracBA instead exhibits smooth, monotonic alignment from shallow to deep layers,
indicating hierarchical integration consistent with multi-scale feature extraction.
This distributed encoding enhances robustness to layer-wise pruning and activation-based defenses.

\begin{figure}[t]
	\centering
	\includegraphics[width=\linewidth]{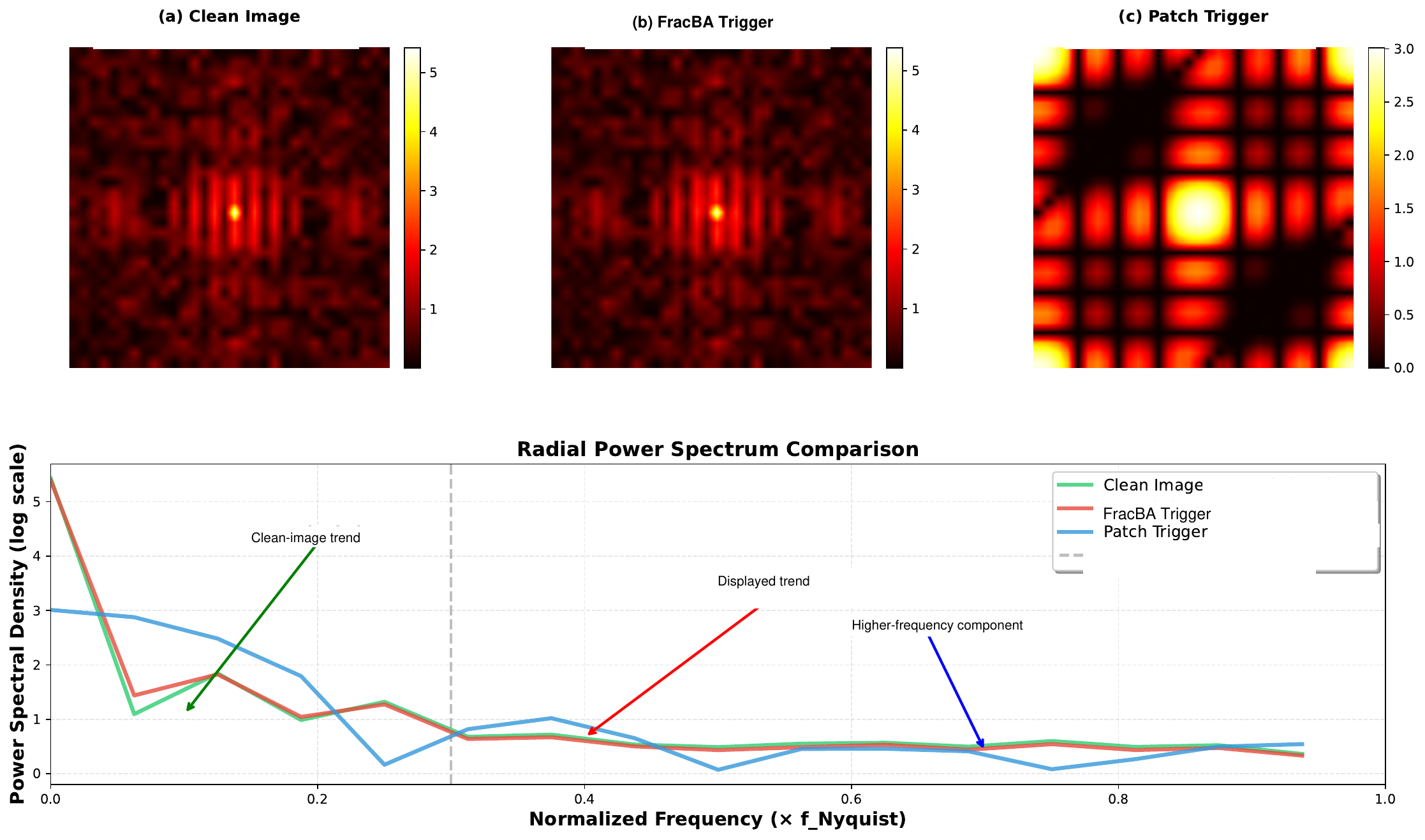}
	\caption{Radial power spectrum comparison of clean, FracBA-triggered, 
		and patch-triggered images. FracBA closely follows natural $1/f^2$ decay, 
		whereas patch triggers introduce sharp high-frequency spikes.}
	\label{fig:frequency}
\end{figure}

\textbf{Frequency Domain Analysis.}
Figure~\ref{fig:frequency} compares radial power spectra.
Patch triggers introduce sharp high-frequency spikes that deviate from natural $1/f^2$ statistics~\cite{zeng2021rethinking}.
FracBA preserves approximate power-law decay with only minor spectral redistribution,
maintaining multi-scale structure while avoiding concentrated frequency anomalies.
This spectral naturalness explains the gradual representational alignment observed across layers.

\begin{figure}[t]
	\centering
	\includegraphics[width=0.48\textwidth]{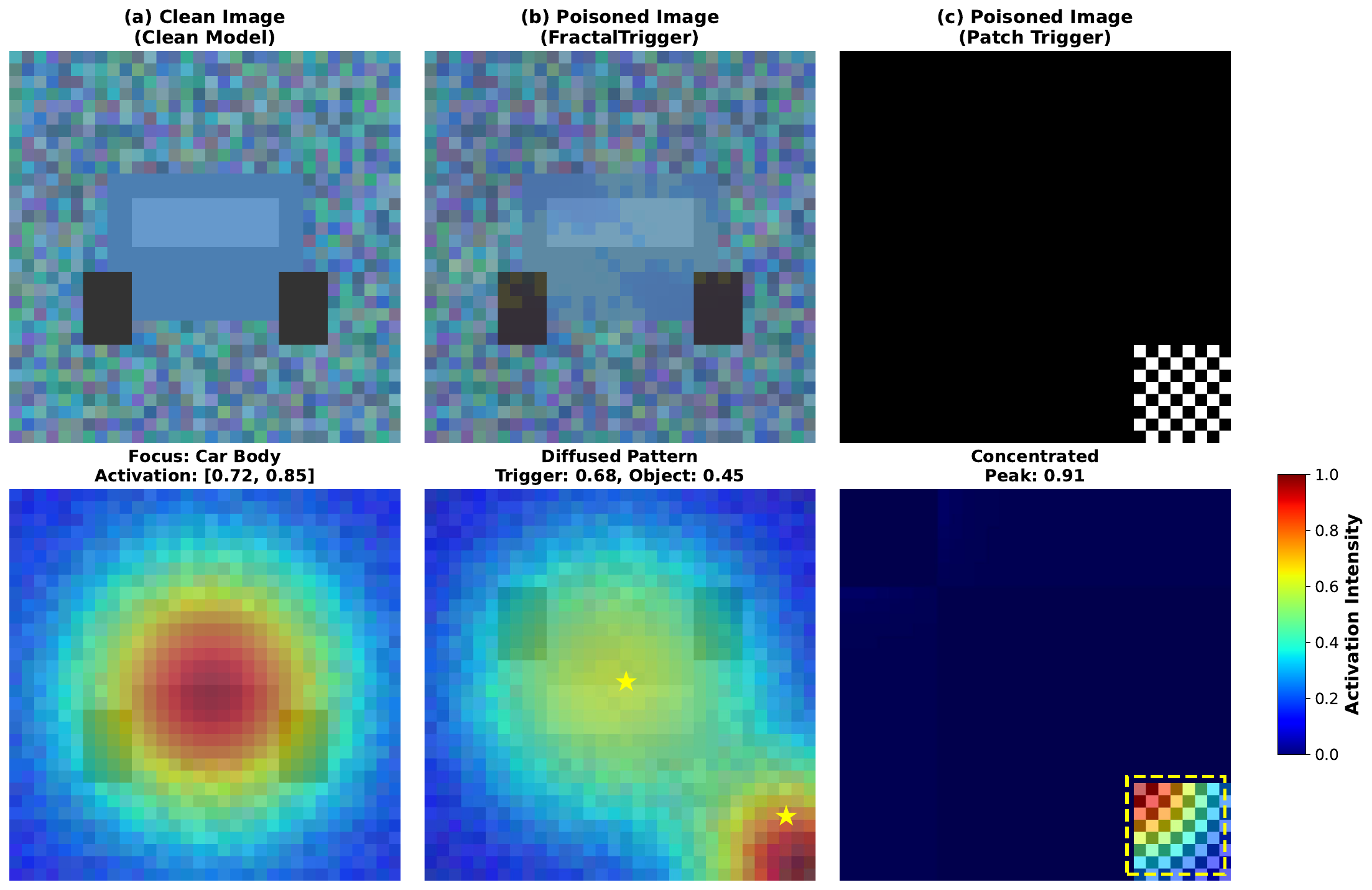}
	\caption{Grad-CAM visualization comparing attention patterns.
		(a) Clean image: attention focuses on the object region.
		(b) FracBA-triggered image: attention spreads across the image, 
		mimicking natural texture processing.
		(c) Patch-triggered image: attention concentrates on a localized peak.}
	\label{fig:gradcam}
\end{figure}

\textbf{Attention and Attribution.}
Grad-CAM visualizations (Figure~\ref{fig:gradcam}) show that patch triggers produce localized attention peaks,
whereas FracBA induces diffused attention resembling natural texture processing.
Integrated gradients further confirm that attribution is distributed across the fractal structure rather than concentrated in a single region.
Such distributed encoding improves resilience against neuron pruning and trigger inversion methods~\cite{nguyen2020input}.

\subsection{Empirical Mechanism Checks}
\label{sec:theory_validation}

We examine whether the limited analytical mechanisms in Section~\ref{sec:theory} are qualitatively consistent with the reported experiments.

\textbf{Gradient Masking Validation.}
The gradient norm constraint suggests that, when the uncalibrated local proxy is close to the benign-update scale, norm-based detection signals may weaken.
Across a wide range of trigger magnitudes and scaling factors not used during model design, the empirical detection behavior reveals three qualitative regimes: low-detection, marginally detectable, and over-scaled.
These results are consistent with the role of update magnitude as one detection signal, but they do not establish a defense-independent detection law.

\textbf{Spectral Property Validation.}
Proposition~\ref{thm:spectral} provides only a generic pointwise feature-stability bound under the Lipschitz condition; the spectral behavior of fractal triggers is evaluated empirically in this section.
Across diverse fractal patterns with varying Hausdorff dimensions, measured spectral slopes are consistent with the expected power-law trend.
Compared with non-fractal filtered-noise controls, spectral naturalness alone reduces detectability but is insufficient for maximal evasion.
Fractal triggers uniquely combine spectral alignment with spatial self-similarity, enabling both low detection rates and high attack success.

\textbf{Iterative versus Spatial Decomposition.}
We conduct an end-to-end ablation under identical experimental conditions to compare the full FracBA configuration with traditional spatial partitioning, as reported in Table~\ref{tab:theory_iter_vs_spatial}. The full FracBA configuration achieves 96.8\% ASR and reaches 90\% ASR in 142$\pm$11 rounds, requiring about 2,600 poisoned samples to reach 80\% ASR. Spatially partitioned DBA achieves 90.2\% ASR, which is re-evaluated under the matched-budget protocol and hence differs slightly from the 89.6\% in the main results table, and reaches 90\% ASR in 168$\pm$14 rounds, requiring about 4,200 poisoned samples. Rotation, scaling, and clipping also differ between these configurations; the comparison therefore does not isolate hierarchical decomposition. The measured direction is consistent with, but does not establish, the proposed structural-coherence explanation.

\begin{table}[t]
\centering
\caption{Controlled comparison of iterative versus spatial decomposition. The FracBA column reports the full default configuration as the reference point; the samples-to-80\%-ASR metric is measured under the matched sample budget for both methods. Note: FLTrust/AC detection rates are measured under the original statistical protocol and are not directly comparable to the local-estimate values in the main row of Table~\ref{tab:defense_resistance}.}
\label{tab:theory_iter_vs_spatial}
\footnotesize
\begin{tabular}{l|cc}
\toprule
\textbf{Metric} & \textbf{Spatial} & \textbf{FracBA} \\
\midrule
ASR (\%) & 90.2$\pm$1.9 & \textbf{96.8$\pm$1.2} \\
Rounds to 90\% ASR & 168$\pm$14 & \textbf{142$\pm$11} \\
Samples to 80\% ASR & 4200 & \textbf{2600} \\
FLTrust detection (\%) & 15.2 & \textbf{6.8} \\
AC detection (\%) & 32.7 & \textbf{17.2} \\
\bottomrule
\end{tabular}
\end{table}

\textbf{Sample-Efficiency Consistency.}
The controlled comparison is directionally consistent with the qualitative mechanism in Section~\ref{sec:theory}. Its finite ASR thresholds and reported protocol do not support extrapolation across client counts, datasets, or data-heterogeneity levels.

\section{Discussion}
\label{sec:discussion}

Our results reveal limitations in the evaluated FL backdoor defenses when confronted with hierarchical fractal triggers. 
Rather than exploiting extreme perturbations or isolated anomalies, FracBA integrates backdoor signals into the natural multi-scale feature hierarchy of neural networks. 
This structural compatibility weakens assumptions underlying many existing defenses.

\subsection{Why Current Defenses Struggle}

Existing defenses rely on three implicit assumptions: 
(1) backdoor triggers induce localized spectral irregularities, 
(2) malicious updates exhibit anomalous gradient magnitudes, and 
(3) backdoor effects manifest within individual training rounds. 

In the reported settings, the observations are consistent with these signals being insufficient to identify this configuration, rather than proving general limitations of the defenses. 
Its fractal structure distributes energy across scales instead of forming concentrated frequency peaks. 
When the local proxy approximates the relevant benign scale, adaptive scaling may reduce separability under magnitude-based filtering. 
Finally, cumulative multi-round injection embeds the backdoor gradually, which may make single-round inspection or post-hoc analysis less informative.

These observations suggest that current defenses are primarily tuned to detect spatially localized or norm-amplified attacks, and are less effective against structurally distributed and temporally accumulated backdoors.

\subsection{Implications for Future Defenses}

Effective defense against fractal-based backdoors likely requires moving beyond cluster detection and norm thresholding. 
Potential directions include multi-scale spectral analysis, structural self-similarity testing, longitudinal monitoring of representation drift across training rounds, and unlearning-based model purification~\cite{li2021anti}. 
More broadly, defense design may need to account for hierarchical and distributed trigger constructions rather than assuming centralized or patch-based patterns.

\subsection{Limitations}

FracBA introduces additional design complexity compared to simple patch-based attacks, as it requires coordinated hierarchical trigger construction across malicious clients. The local norm proxy used for scaling has not been calibrated to the benign-client population under heterogeneous data; the associated defense results should therefore be interpreted as configuration-specific.
Its advantages also diminish when very few malicious clients actually participate per round (Table~\ref{tab:ablation_sparse}: $k=2$, ASR drops to 76.2\%), reducing the partial-reconstruction benefit of hierarchical decomposition; note that this axis differs from the malicious-ratio axis in Table~\ref{tab:ablation_malicious}, where FracBA still attains 84.6\% ASR at a 5\% malicious ratio. These trade-offs highlight the tension between structural sophistication and operational simplicity, and motivate future exploration of adaptive or simplified hierarchical designs. In addition, our evaluation focuses on natural-image benchmarks; performance on medical imaging, high-resolution heterogeneous images, and constrained deployment conditions has not yet been validated, and we leave these settings for future work.

\section{Conclusion}

We proposed FracBA, a hierarchical backdoor attack for FL that leverages fractal geometry and distributed generation. 
By decomposing the trigger generation process across clients rather than partitioning a completed pattern, FracBA achieves high ASR in the evaluated FedAvg settings, while the reported end-to-end comparison with spatial DBA shows a lower poisoned-sample requirement at the stated threshold. This comparison does not isolate hierarchical decomposition from rotation, scaling, or clipping.

The effectiveness of FracBA arises from three interacting principles: 
(1) multi-scale fractal structure that aligns with hierarchical feature extraction, 
(2) adaptive gradient scaling that preserves statistical consistency with benign updates, and 
(3) cumulative multi-round injection that embeds backdoor signals gradually. 
Together, these mechanisms indicate potential structural blind spots in current federated defenses, particularly those relying on localized spectral anomalies or norm-based filtering.

Our findings suggest that defending against distributed and hierarchical backdoor attacks requires moving beyond spatial or magnitude-based assumptions toward multi-scale and temporally aware analysis. 
We hope this work stimulates further research on structurally informed defenses for FL.

\bibliographystyle{IEEEtran}
\bibliography{references}

\end{document}